\documentclass{jfm}

\usepackage{graphicx}
\usepackage{newtxtext}
\usepackage{newtxmath}
\usepackage{natbib}

\usepackage{booktabs}  
\usepackage{siunitx} 

\usepackage{hyperref}
\hypersetup{
    colorlinks = true,
    urlcolor   = blue,
    citecolor  = black,
}

\newcommand{\RomanNumeralCaps}[1]

\title{Reynolds number effects on surface-induced secondary flows in turbulent boundary layers}

\author{T. Medjnoun
  \corresp{\email{t.medjnoun@soton.ac.uk}},
  M. Nillson-Takeuchi
 \and B. Ganapathisubramani}

\affiliation{Department of Aeronautics and Astronautics, University of Southampton SO17 1BJ, United Kingdom}

\begin{document}
\maketitle

\begin{abstract}
This study explores the effect of friction Reynolds number ($Re_\tau \approx 3{,}000$--$13{,}000$) on secondary flows in three-dimensional turbulent boundary layers induced by spanwise surface heterogeneity. Using a combination of floating-element drag balance and high-resolution hot-wire anemometry, we examine how varying spanwise spacing ($S/\delta$) influences frictional drag, turbulence intensity, spectral energy distribution, and the organisation of coherent structures. The results reveal that secondary flows modulate turbulence differently depending on $S/\delta$, with strong near-wall effects at $S/\delta < 1$ and outer-layer modulation at $S/\delta \gtrsim 1$. A robust spectral signature of secondary flows peaking at $\lambda_x \approx 3\delta$ and $y \approx 0.5\delta$ emerges across all cases. This peak coexists with, or suppresses, very-large-scale motions (VLSMs), depending on flow region and spacing. While VLSMs are suppressed in low-momentum pathways (LMPs), they gradually recover in high-momentum pathways (HMPs) at higher $S/\delta$ and $Re_\tau$. These findings offer new insight into the interplay between secondary motions and scale interactions in three-dimensional turbulent boundary layers, with implications for drag control, mixing, and surface design.
\end{abstract}



\section{Introduction}

Over the past century, the study of rough-wall turbulence continues to challenge researchers, reflecting the profound complexities of this field and its critical implications across a wide range of engineering applications. Central to these investigations is the ongoing challenge of accurately characterising drag as a function of surface properties, particularly at full-scale conditions. While significant progress has been achieved through advancements in experimental and numerical capabilities, key foundational questions remain unanswered, requiring further targeted research \citep{chung2021}.

Predictive frameworks of frictional drag often assume homogeneous surface roughness. However, this assumption fails for surfaces characterised by spanwise topographical heterogeneity, a feature common in many real-world applications. Examples include environmental flows, such as fluvial sediment transport and bedform formation \citep{Chiu1966a,Colombini1993,McLelland1999,Nikora2012a,Scherer2022}, atmospheric boundary layer flows over wind farms and heterogeneous terrains \citep{markfort2012turbulent,stevens2016effects,alfredsson2017introduction,bossuyt2018effect,Medjnoun2021}, and engineering applications like renewable energy systems and heat exchangers \citep{barros2014,chitrakar2016study,Pathikonda2017,Deyn2022a,wang2024characterization}, as well as marine transportation systems \citep{Monty2016a,Murphy2018,Nugroho2021,kaminaris2023secondary,Medjnoun2023}. Under specific conditions \textendash commonly defined by the relative topography/roughness height/strength and spanwise characteristic length scale \textendash such heterogeneous surfaces can promote flow channelling \citep{MejiaAlvarez2013}, resulting in high- and low-momentum pathways (HMPs and LMPs, respectively). These features are associated with large-scale secondary flows, known as Prandtl’s secondary flows of the second kind, which are critical to understanding the interplay between surface topology and turbulent flow dynamics \citep{Hinze1973,Anderson2015,Kevin2017,Hwang2018,Stroh2020}.

Despite significant advances, critical gaps remain in understanding drag over spanwise heterogeneous surfaces, particularly at high Reynolds numbers. Current research predominantly focuses on low to moderate Reynolds numbers \citep{Medjnoun2018,Nikora2019,Nugroho2021,guo2022energy,Deyn2022,Frohnapfel2024}, often overlooking turbulence dynamics at higher Reynolds numbers where the scale separation between near-wall and outer-layer motions becomes more pronounced \citep{Smits2011a}. Addressing these gaps requires experimental investigations that bridge fundamental understanding and practical relevance at high Reynolds numbers. Such data are essential for developing predictive frameworks to manage drag in engineering systems, such as biofouled ship hulls, and to improve sediment transport models in environmental applications.

Furthermore, secondary flows have the potential to play a significant role in influencing the naturally occurring coherent structures characteristic of wall-bounded turbulence. These structures, spanning scales from near-wall motions to large-scale outer-layer motions, are well described by the attached eddy hypothesis \citep{Townsend1976,Perry1990}, which posits a self-similar hierarchy of eddies scaling with their distance from the wall. The near-wall cycle \textendash a self-sustaining process involving the generation of streamwise streaks and energy production that sustains larger flow structures \textendash forms the foundational energy-generating mechanism in wall-bounded turbulence \citep{Jimenez1999}. \citet{Marusic2019} highlight the hierarchical organisation of attached eddies in wall-bounded turbulence and their role as energy and momentum carriers across scales. At high Reynolds numbers, interactions between inner- and outer-layer motions become increasingly significant as large outer-layer eddies penetrate the near-wall region, modulating local structures and amplifying turbulence intensity within the boundary layer \citep{Hutchins2007,Hutchins2007a}. These dynamics ultimately coalesce into VLSMs, sometimes also referred to as ``superstructures'', which act as conveyors of energy and momentum across the boundary layer \citep{Marusic2010,Smits2011a}. The streamwise coherence and scale of VLSMs, which can exceed an order of magnitude of the boundary layer thickness, highlight their critical role in maintaining energy transfer and turbulence dynamics \citep{Guala2006}.

The exact nature of VLSMs remains debated, particularly regarding their independence and interaction with other scales. Traditionally, VLSMs are considered to form through the alignment or aggregation of smaller Large-Scale Motions (LSMs). In this view, VLSMs are not entirely independent but arise from the spatial organisation of these smaller eddies into large-scale coherent motions \citep{Kim1999,Adrian2000,Deshpande2023}. Alternatively, some studies suggest that VLSMs may behave as independent, self-sustaining structures with distinct dynamical properties, particularly at high Reynolds numbers where scale separation is more pronounced \citep{Guala2006,Jimenez2012,Zampiron2024}. These two interpretations have significant implications for how VLSMs interact with secondary flows. If VLSMs are independent, secondary flows may directly influence their energy and structure. Conversely, if VLSMs primarily arise from aligned LSMs, secondary flows might affect VLSMs indirectly by modulating the underlying LSMs. Understanding this distinction is critical to characterising turbulence at multiple scales and its sensitivity to external influences like surface heterogeneity.

Early insights into these interactions emerged from \citet{Nugroho2013}, who introduced ``herringbone'' riblet-patterned surfaces to passively control turbulent boundary layers and perturb the naturally occurring large-scale coherent motions. Their study revealed substantial modifications in the energy distribution among the largest energetic structures across the boundary layer. Subsequent works \citep{Pathikonda2017,Awasthi2018} examined amplitude and frequency modulation effects for irregular and regular roughness arrangements, respectively. \citet{Pathikonda2017} reported that roughness enhances amplitude and frequency modulations close to the wall, while \citet{Awasthi2018} demonstrated that a distinct outer peak associated with large-scale motions, termed ``modulators'', is preserved within LMPs but vanishes in HMP zones, where turbulent coherent motions exhibit steeper and shorter spatial coherence. 

Additional studies have explored different types of surface heterogeneity and their effects on the spectral and structural attributes of various flows. For instance, in their experimental study of a turbulent boundary layer over a ridge-type surface, \citet{Medjnoun2018} revealed changes in energy redistribution across all scales, with effects varying according to the spanwise wavelength of the surface heterogeneity. In a direct numerical simulation of a fully turbulent pipe flow over three-dimensional wavy topography, \citet{Chan2018} showed that energy is reorganised from the largest to the shorter streamwise wavelengths, with the spectral peaks correlated with the characteristic roughness length scale. Using a similar surface roughness to that of \citet{Pathikonda2017}, \citet{Barros2019} found significant spectral content variations across the spanwise direction of the surface heterogeneity. More specifically, a shift in energy content to shorter wavelengths above the LMPs, whereas the impact was less pronounced in the HMPs

More recently, in their studies of open-channel flows, \citet{Zampiron2020} and \citet{Luo2023} demonstrated that for spanwise spacings $S/\delta \leq 1$, VLSMs are suppressed, arguing that secondary motions absorb VLSM energy. They also identified a signature of secondary flow instability, manifesting as the meandering of HMPs and LMPs. Similarly, using large-eddy simulations of turbulent flow over a randomly rough surface, \citet{Ma2023} found that secondary motions play a crucial role in redistributing energy, enhancing turbulence intensity in the outer region, and modifying spectral peaks. Their results further suggest that roughness-induced secondary motions alter the organisation of coherent structures, leading to deviations from classical outer-layer similarity.

Consistent with \citet{Zampiron2020}, \citet{Wangsawijaya2020} observed the meandering behaviour of secondary motions and later characterised their unsteadiness in a strip-type heterogeneous rough-wall boundary layer \citep{Wangsawijaya2022}. They found that secondary motions and large-scale structures can coexist within certain limits, with maximum secondary flow instabilities occurring at $S/\delta \sim \mathcal{O}(1)$. More recently, \citet{Zhdanov2024} numerically investigated the influence of ridge-type heterogeneity in turbulent channel flow and reported a significant increase in energy content at higher wavelengths at spanwise locations corresponding to the centres of secondary flows. While these studies provide valuable insights, they primarily focus on low to moderate Reynolds numbers, leaving the dynamics at high Reynolds numbers largely unexplored.

Building on this foundation, this study investigates the interaction between ridge-induced secondary flows and turbulent boundary layers at high Reynolds numbers. Using an in-house floating-element drag balance (FEDB) for wall-shear stress measurements \textendash a direct method avoiding any flow assumptions \textendash and hot-wire anemometry (HWA) for spectral analysis, we address critical gaps in understanding drag behaviour, energy redistribution, and turbulent structure modulation over spanwise heterogeneous surfaces across different Reynolds numbers.

\section{Experimental Methods}
Experiments were conducted in the newly commissioned Boundary Layer Wind Tunnel (BLWT), a closed-return facility at the University of Southampton. This Göttingen-type tunnel has a 12\,m long test section in the \( x \)-direction, with a cross-sectional area of 1\,m \(\times\) 1.2\,m in the wall-normal (\( y \)-direction) and spanwise (\( z \)-direction) directions. The BLWT features a 7:1 contraction in the flow-conditioning section to ensure a uniform inflow profile, with freestream turbulence intensity consistently below 0.1\% within the operational speed range. A cooling unit maintains the air temperature within \(\pm0.1\,^{\circ}\mathrm{C}\), ensuring stability for temperature-sensitive measurements. This precise temperature control is achieved using two heat exchangers and a PID temperature controller.

\begin{figure}
\centering
\includegraphics[width= \textwidth]{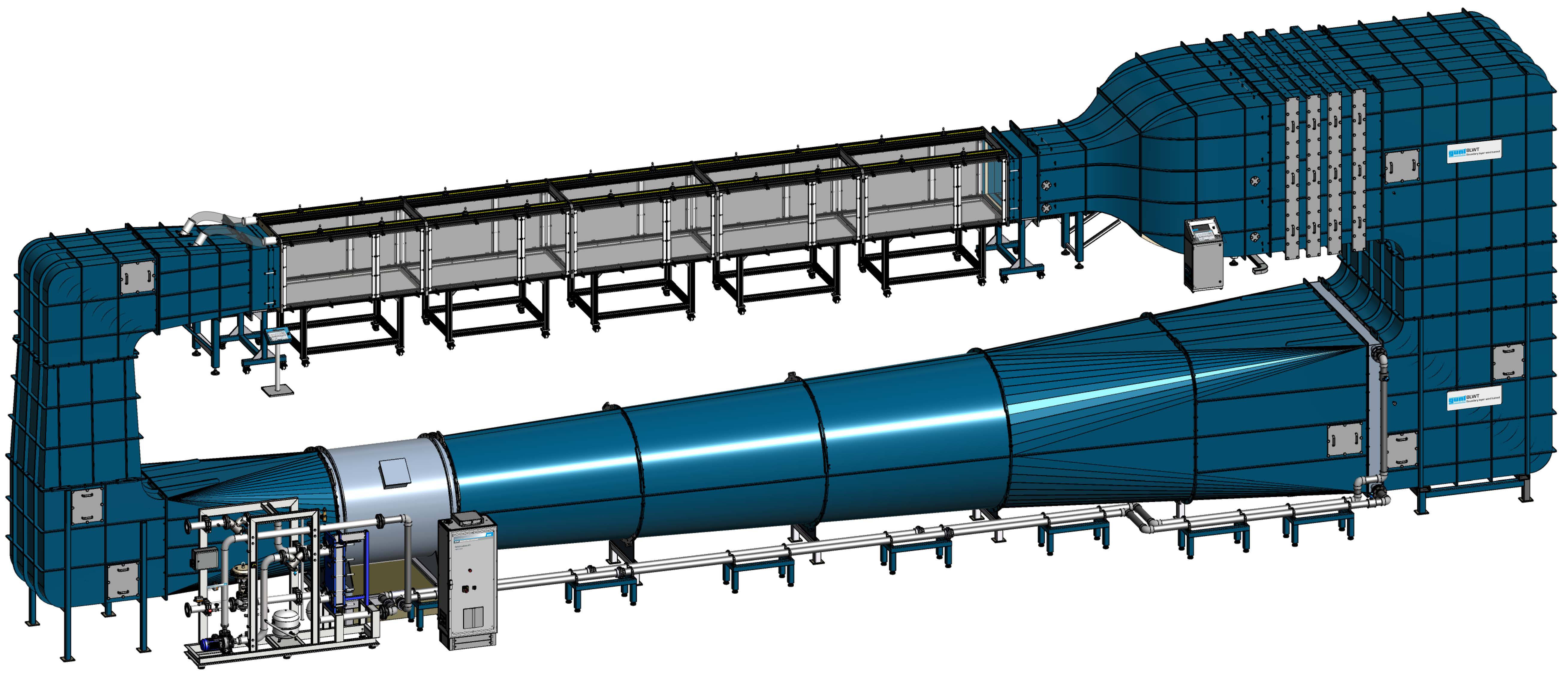}

\includegraphics[width= 0.8\textwidth]{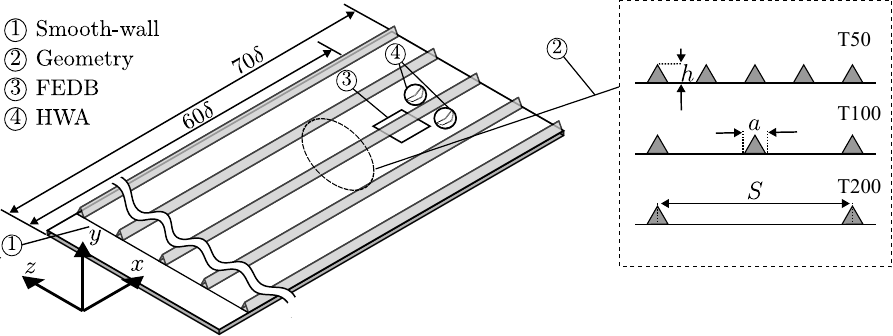}
\caption{(top) Schematic of the boundary layer wind tunnel (BLWT) at the University of Southampton; (bottom) surface arrangement showing the spanwise-heterogeneous ridge-type surface, including an illustration of the experimental methods.}
\label{figure1}
\end{figure}

Boundary layers develop directly on the wind tunnel floor under a slightly favourable pressure gradient due to the fixed cross-sectional area of the test section. To ensure a laminar-to-turbulent transition, the boundary layer was tripped 50\,mm from the leading edge of the test section using a zig-zag turbulator tape with a 6\,mm point-to-point spacing and a thickness of 0.5\,mm. At the measurement location (approximately $60\delta$ from the inlet), the floorboard includes a central cutout to accommodate the FEDB. Static-pressure taps were installed around the cutout, approximately 50\,mm from its centre, to measure the local streamwise pressure gradient and apply corrections to the FEDB measurements. Figure~\ref{figure1} provides a schematic of the BLWT, detailing the experimental methods and surface configurations employed.

\subsection{Surface Heterogeneity}
Surface heterogeneity was introduced by affixing longitudinal triangular ridges to a smooth base. These ridges, modelled on the HS2 configuration by \cite{Medjnoun2020}, are equilateral triangles with a side length of $a = 6.4$\,mm and a height of $h = 5.6$\,mm. Three spanwise spacings ($S$) were tested: $S = 50$\,mm, $100$\,mm, and $200$\,mm, corresponding to configurations labelled T50, T100, and T200, respectively. These spacings are equivalent to $S/\delta \approx 0.3, 0.6$, and $1.3$, with $h/\delta \approx 4\%$, when scaled with the spanwise-averaged boundary layer thickness $\delta$.
\subsection{Drag Balance}
Wall shear stress was measured using an in-house floating-element drag balance (FEDB) designed with a zero-displacement force-feedback system \citep{AguiarFerreira2024}. The floating element, a square of 200\,mm side length, was installed flush with the wind tunnel floor, located approximately $60\delta$ from the inlet. The FEDB was subjected to nine free-stream velocities ($U_{\infty}$) ranging from 10 to 45\,m/s to determine the frictional drag coefficient $C_f$ as a function of the Reynolds number. Each acquisition lasted 120 seconds, sampled at 256\,Hz, capturing at least 2,500 boundary-layer eddy turnover times ($\tau_{eddy} = \delta/U_{\infty}$) for the lowest velocity. Five repetitions were conducted at each velocity to ensure statistical significance. Pre- and post-calibration procedures were performed for each configuration using a standard force transducer. The calibration coefficient varied by less than 0.5\%, ensuring high precision and negligible drift over extended measurements. The overall uncertainty in the skin-friction coefficient (precision and systematic uncertainties) for all the cases was less than $\pm 1.5 \%$. A more detailed and comprehensive discussion of the FEDB design, acquisition procedures, and measurement uncertainties is available in \citet{AguiarFerreira2024}.
\subsection{Hot-Wire Anemometry}
Time-resolved velocity measurements were conducted using two single Auspex A55P05 boundary-layer hot-wire probes. These probes consist of tungsten wires with a diameter of 5\,$\mu$m and a sensing length of 1\,mm, yielding a length-to-diameter ratio of 200, in accordance with the recommendations of \citet{Hutchins2009a}. The inner-scaled length of the HWA sensor $L^{+}=LU_{\tau}/\nu$ ranged between 25 and 100 viscous length scales. Measurements were recorded simultaneously at two symmetry planes: $z/S = 0$ (valley symmetry plane) and $z/S = 0.5$ (ridge symmetry plane).

The setup was positioned at the same location as the FEDB, and measurements were performed at three freestream velocities: $U_{\infty} \approx 10$, $20$, and $40$\,m/s. A DANTEC Streamline Pro Constant Temperature Anemometer (CTA) system was employed, operating at a fixed overheat ratio of 0.8. The wall-normal flow was traversed logarithmically at 50 locations, from $0.002\delta$ to $2\delta$. Data were recorded for 3 to 10 minutes at each position, capturing a minimum of 25,000 boundary layer eddy turnover cycles, sufficient for convergence of turbulence intensity and spectral measurements \citep{Hutchins2009a}.

The raw signal was amplified, low-pass filtered at 30\,kHz, and sampled at 60\,kHz using a 16-bit National Instruments NI-DAQ USB 6212 system. Freestream velocity was monitored using a Pitot-static probe connected to a Furness Controls FCO560 micromanometer. Additionally, a Dantec 90P10 temperature probe measured air temperature, correcting for thermal drift. Pre- and post-calibration procedures were conducted, and time-based interpolation between calibration coefficients accounted for temperature and electrical drifts, with an overall drift of less than 1\%.

\section{Results}\label{sec:results}

This section presents the analysis of results obtained from the FEDB and HWA measurements. Section \ref{sec:drag} investigates the behaviour of frictional drag at high Reynolds numbers under the influence of spanwise surface heterogeneity. Section \ref{sec:FlowHet} explores the effects of Reynolds number on flow heterogeneity. Sections \ref{sec:TurbInt}, \ref{sec:Spectra}, and \ref{sec:VLSMs} examine the influence of surface heterogeneity and Reynolds number on turbulence intensity, spectral characteristics, and the VLSMs, respectively.

\subsection{Frictional drag}\label{sec:drag}

\begin{table}
    \centering
    \sisetup{table-format=1.3, exponent-product=\times} 
    \begin{tabular}{S[table-format=2.0] S[table-format=1.2] *{8}{S[table-format=1.3]}}
         & & 
        \multicolumn{2}{c}{T50 (\(\blacktriangle\))} & 
        \multicolumn{2}{c}{T100 (\(\blacksquare\))} & 
        \multicolumn{2}{c}{T200 (\(\blacklozenge\))} & 
        \multicolumn{2}{c}{Smooth (\(\bullet\))} \\
        \cmidrule(lr){3-4} \cmidrule(lr){5-6} \cmidrule(lr){7-8} \cmidrule(lr){9-10}
        \multicolumn{1}{c}{\( U_\infty \) (m/s)} &
        \multicolumn{1}{c}{\( Re_x \) (\(\times 10^5\))} & 
        \multicolumn{1}{c}{\( C_f \) (\(\times 10^3\))} & 
        \multicolumn{1}{c}{\( \Delta U^{+} \)}  &
        \multicolumn{1}{c}{\( C_f \) (\(\times 10^3\))} &
        \multicolumn{1}{c}{\( \Delta U^{+} \)}  &
        \multicolumn{1}{c}{\( C_f \) (\(\times 10^3\))} &
        \multicolumn{1}{c}{\( \Delta U^{+} \)} &
        \multicolumn{1}{c}{\( C_f \) (\(\times 10^3\))} &
        \multicolumn{1}{c}{\( \Delta U^{+} \)} \\        
        \midrule
        \textcolor{blue}{10}  & \textcolor{blue}{0.60} & 2.96 & 2.44 & 3.06 & 2.90 & 3.00 & 2.59 & 2.81 & \textendash \\
        15  & 0.90 & 2.81 & 2.71 & 2.84 & 2.84 & 2.73 & 2.30 & 2.49 & \textendash\\
        \textcolor{blue}{20}  & \textcolor{blue}{1.20} & 2.70 & 2.88 & 2.70 & 2.84 & 2.57 & 2.15 & 2.35 & \textendash\\
        25  & 1.50 & 2.62 & 3.02 & 2.60 & 2.85 & 2.47 & 2.09 & 2.27 & \textendash\\
        30  & 1.81 & 2.56 & 3.13 & 2.53 & 2.90 & 2.40 & 2.05 & 2.21 & \textendash\\
        35  & 2.11 & 2.52 & 3.24 & 2.48 & 2.99 & 2.33 & 2.05 & 2.15 & \textendash\\
        \textcolor{blue}{40}  & \textcolor{blue}{2.42} & 2.48 & 3.35 & 2.43 & 3.04 & 2.28 & 2.01 & 2.11 & \textendash\\
        45  & 2.69 & 2.48 & 3.66 & 2.41 & 3.16 & 2.24 & 2.01 & 2.07 & \textendash\\
        \bottomrule
    \end{tabular}
    \caption{Skin friction coefficient \( C_f \) and its associated roughness function \( \Delta U^{+} \) for the different test cases and Reynolds numbers \( Re_{x}\). The blue-highlighted cases represent flow conditions where HWA measurements are conducted.}
    \label{tab:drag_results}
\end{table}

\begin{figure}
\includegraphics[width=0.45\textwidth]{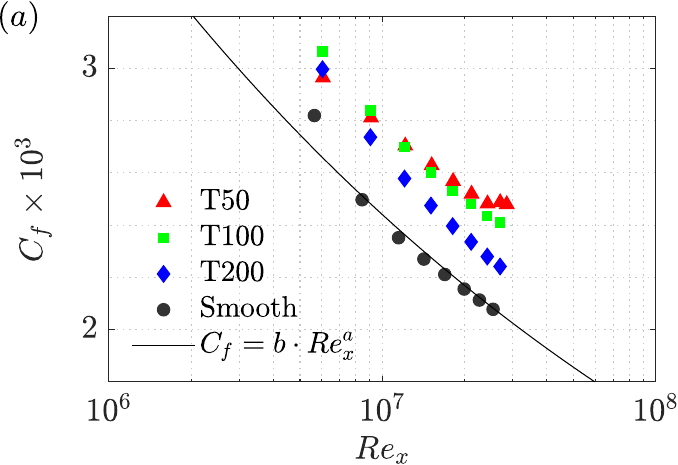}
\includegraphics[width=0.45\textwidth]{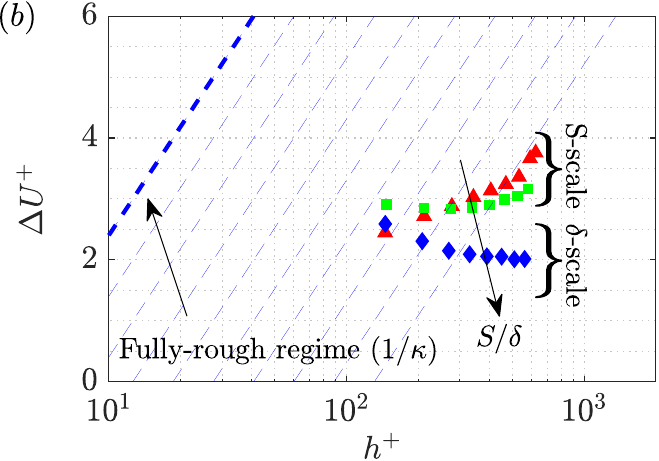}
\caption{($a$) Variation of the skin-friction coefficient as a function of Reynolds number, compared with the smooth-wall baseline and Schlichting power law, and ($b$) the associated roughness function. The blue dashed line represents the classical ’homogeneous’ fully-rough regime, with a $1/\kappa$ slope.}
\label{figure2}
\end{figure}

The results from the FEDB are illustrated in figure~\ref{figure2}, depicting the response of the wall shear stress to changes in the spanwise spacings of the surface heterogeneity. Figure~\ref{figure2}($a$) illustrates the variation of the skin-friction coefficient $C_{f}$ as a function of $Re_{x}$. At low Reynolds numbers, there is only a weak difference in magnitude between the three surfaces. However, as $Re_{x}$ increases, a proportional difference begins to manifest (around $Re_{x} \approx 10^7$), with T200 showing a noticeable reduction in magnitude compared to T50 and T100. 

At the highest Reynolds number measured (approximately $Re_{x} \approx 3 \times 10^7$), T50 exhibits the highest frictional drag, followed by T100 and then T200. This outcome is expected since T50 has a larger surface area compared to T100 and T200. Given $C_{f}$ scales proportionally with the planform solidity, $C_{f}$ naturally increases as the spanwise wavelength decreases, and vice versa. However, regardless of the spanwise spacing between ridges, the drag remains higher than that of the smooth wall, as shown by the 25$\%$ higher value for T50 at the largest $Re_{x}$, in virtue of the planform solidity argument.

Figure~\ref{figure2}($a$) also reveals differences in the decay rate of $C_{f}$ across cases. The skin-friction coefficient for T50 appears to reach an onset of an asymptote at high $Re_{x}$, as seen in the nearly constant values at the last measurement points. In contrast, T200 decays similarly to the smooth wall flow but maintains a relatively constant offset in frictional drag. Additionally, the difference (relative increase) in frictional drag as the spanwise wavelength is reduced (from T200 to T50) becomes marginally smaller at higher Reynolds numbers. This raises the possibility that for even smaller wavelengths, the skin-friction coefficient may asymptote at an earlier Reynolds number. Conversely, for larger wavelengths, the frictional drag asymptotes towards the smooth-wall curve. Together, these trends delineate a bounding area of expected frictional drag for a given geometry, varying between homogeneous-rough-like behavior for smaller wavelengths and homogeneous-smooth-like decay for large wavelengths.

The differences in frictional drag between the rough and smooth surfaces can be examined by quantifying the roughness function $\Delta U^+$ at matched frictional Reynolds numbers ($Re_{\tau}$). Assuming the difference in the wake strength parameter $\Pi$ is invariant with Reynolds number, the roughness function can be expressed as:
\begin{equation}
\Delta U^+ =  \sqrt{\frac{2}{C^{S}_f}} - \sqrt{\frac{2}{C^{R}_f}}.
\label{dup}
\end{equation}

Figure~\ref{figure2}($b$) shows the roughness function results. The values of $\Delta U^+$ remain relatively low, ranging from 2 to 4 at the highest Reynolds numbers, which is significantly lower than typical rough-wall flows at equivalent Reynolds numbers (\citealt{Schultz2007,Castro2013,Flack2014,Squire2016,Medjnoun2023}). It is well-established that Reynolds number invariance in $C_{f}$ primarily arises from pressure drag contributions, which dominate at sufficiently high Reynolds numbers, leading to the fully rough regime (\citealp{Napoli2008,Yuan2014}). However, such contributions require wake-producing roughness features (i.e., streamwise flow separation), which are absent in these spanwise heterogeneous surfaces. Therefore, the low $\Delta U^+$ values corroborate the absence of pressure-drag-producing roughness elements. This observation aligns with recent studies that report similar trends in heterogeneous surfaces lacking wake-producing roughness features or exhibiting only weak wake formation (\citealp{Medjnoun2018,Medjnoun2020,Nugroho2021,Deyn2022,Frohnapfel2024}).

Despite the low magnitude of $\Delta U^+$, the variation of $\Delta U^{+}(h^{+})$ for T50 (and to a lesser extent, T100) approaches the $1/\kappa$ asymptote, suggesting the possible emergence of an aerodynamic roughness length scale ($h_{s}$) at either higher Reynolds numbers or smaller spanwise wavelengths (i.e., $S$-scale surfaces). In contrast, for larger wavelengths (T200), $\Delta U^+$ remains invariant with $h^+$, consistent with figure~\ref{figure2}($a$). These results indicate that frictional drag for smooth spanwise heterogeneous surfaces follows two asymptotic limits, a bounded envelope between rough-like and smooth-wall-like behaviors, depending on the relative scale of $S$ to $\delta$. When $S/\delta \ll 1$, flow heterogeneity and secondary flow significance depend on $S$, and the roughness function exhibits $k$-type behavior for small spanwise wavelengths. Conversely, when $S/\delta \geq 1$, flow heterogeneity is primarily governed by $\delta$, with behavior asymptoting toward the smooth-wall regime as spanwise wavelengths increase.

These findings suggest that secondary flows play a crucial role in redistributing momentum and energy across the turbulent boundary layer. For smaller spanwise wavelengths (e.g., T50), near-wall-confined secondary flows significantly increase wall-shear stress, leading to enhanced drag. In contrast, larger wavelengths (e.g., T200) generate large-scale secondary flows that predominantly affect the outer region of the boundary layer, exerting a reduced influence on the near-wall region. This results in drag behaviour closely resembling that of smooth-wall conditions. It is important to note that these results are currently applicable only to smooth spanwise heterogeneous surfaces. Further investigation is required to understand the behaviour of pressure-drag-producing spanwise heterogeneous surfaces, which are expected to behave differently, as recently demonstrated by \citet{Medjnoun2021} and \citet{Frohnapfel2024}.

\subsection{Flow heterogeneity}\label{sec:FlowHet}

\begin{figure}
\centering
\includegraphics[width=0.48\textwidth]{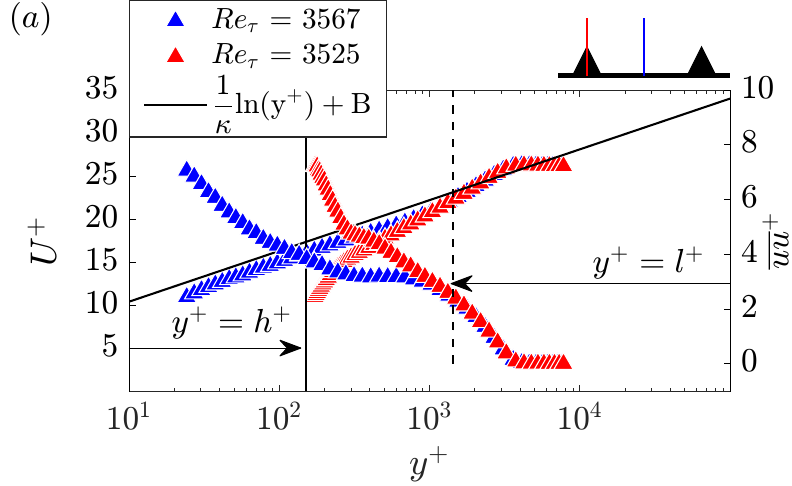}
\includegraphics[width=0.48\textwidth]{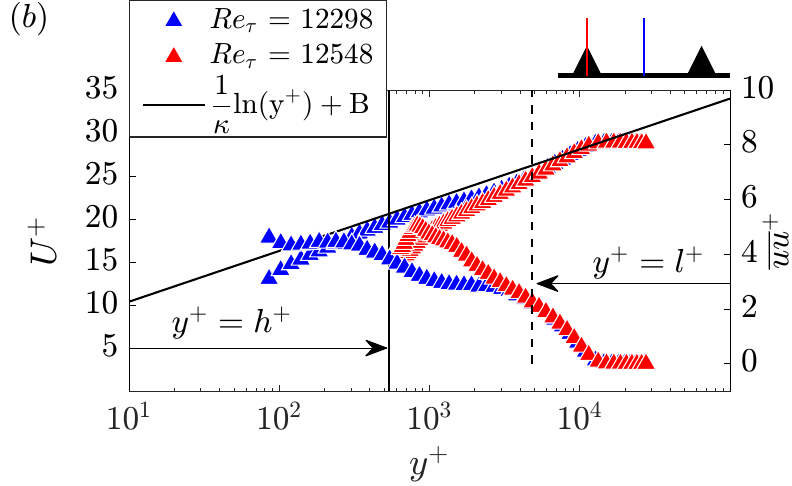}
\caption{Inner-scaled mean streamwise velocity and variance profiles above the ridge (red) and valley (blue) at ($a$) low and ($b$) high Reynolds number for the T50 case. The log-law slope (solid black line) is represented with constants taken as 0.39 and 4.3 for $\kappa$ and $B$, respectively. The vertical solid line shows the inner-normalised ridge height ($y^{+}=h^{+}$) whereas the vertical dashed line ($y^{+}=l^{+}$) represents the wall-normal extent of the spanwise mean and turbulence flow heterogeneity i.e. the height at which the $U^{+}_{valley}=U^{+}_{ridge}$ and $\overline{uu}^{+}_{valley}=\overline{uu}^{+}_{ridge}$.}
\label{figure3}
\end{figure}

To assess the effect of surface condition on turbulent boundary layer flow heterogeneity, both the ridge (red) and valley (blue) symmetry planes ($z/S = 0, 0.5$) are illustrated in figure~\ref{figure3}. The mean streamwise velocity ($U^+$) and turbulence intensity ($\overline{uu}^{+}$) profiles are normalised by the spanwise wavelength-averaged friction velocity $U_{\tau}$, measured using the FEDB. The turbulence variance profiles have been corrected for attenuation due to unresolved contributions in the streamwise stress, following \cite{Smits2011}.

Figure~\ref{figure3} presents results for the T50 case at low ($Re_\tau \approx 3500$) and high ($Re_\tau \approx 12500$) Reynolds numbers, with similar trends observed across other cases. Using area-averaged friction velocity for normalisation, the profiles reveal a horizontal shift due to the ridge height ($h^+$), with the ridge profile shifted toward higher $y^+$ compared to the valley. This horizontal offset is accompanied by a relatively small vertical shift in $U^+$ but a pronounced difference in $\overline{uu}^{+}$. 

The valley (blue) profiles exhibit higher $U^+$, corresponding to high-momentum pathways (HMPs), while the ridges (red) form low-momentum pathways (LMPs). This behavior aligns with previous studies that examined ridge-type spanwise heterogeneous surfaces (\citealp{Nezu1993, Wang2006, Vanderwel2015, Hwang2018,Castro2021,Zhdanov2024}). The turbulence intensity is redistributed, with LMPs exhibiting elevated $\overline{uu}^{+}$, while HMPs show reduced turbulence. These observations reflect the ability of large-scale secondary motions to draw energy from valleys and redistribute it toward ridges through sustained upwash and downwash mechanisms \citep{Anderson2015}.

\begin{figure}
\centering
\includegraphics[width=0.35\textwidth]{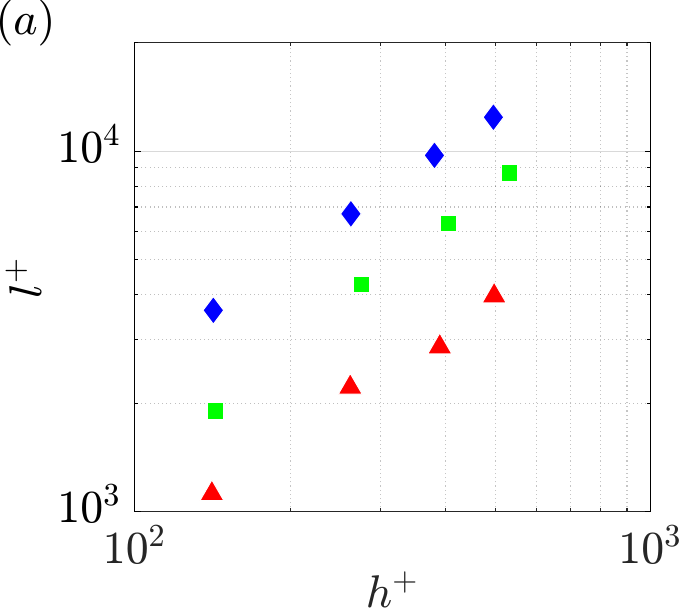}
\includegraphics[width=0.35\textwidth]{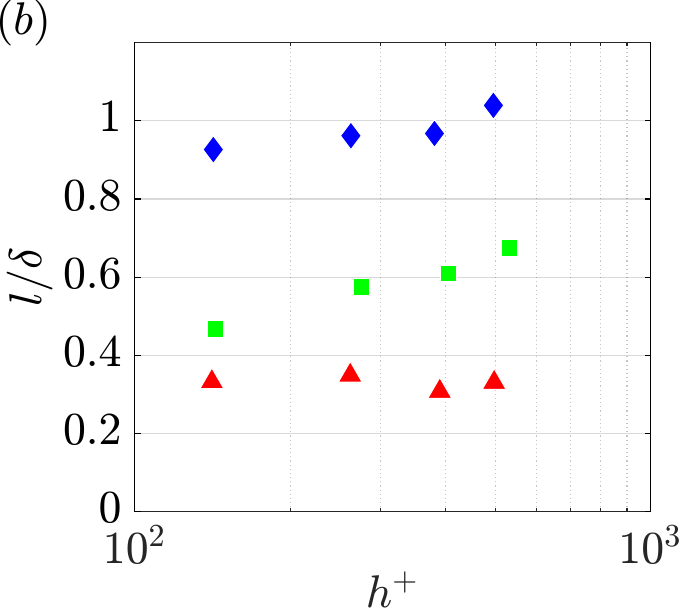}
\caption{Variation of the wall-normal extent of the turbulence heterogeneity normalised in ($a$) inner and ($b$) outer units, as a function of the inner-normalised ridge height, for the different cases and Reynolds numbers.}
\label{figure4}
\end{figure}

At high Reynolds numbers, the turbulence variance profiles reveal additional energetic regions. Specifically, the ridge profile shows a second energetic plateau near $y^+ \approx 1000$, suggesting strong interactions between secondary flows and outer-layer turbulence. This plateau marks a significant departure from low Reynolds number behavior, where turbulence is primarily confined near the wall.

Farther from the wall, both $U^+$ and $\overline{uu}^{+}$ profiles collapse and asymptote toward $U_\infty^+$ and $0$, respectively. This merging point, denoted $l^+$, represents the height at which spanwise homogeneity is restored, marking the extent of secondary flow influence.

Figure~\ref{figure4} further examines the scaling of $l^+$ with $h^+$ in inner units (figure~\ref{figure4}($a$)) and with $\delta$ in outer units (figure~\ref{figure4}($b$)). The results reveal a linear relationship between $l^+$ and $h^+$ in log-log scaling, indicating a strong but expected dependence of flow heterogeneity on local geometry. However, when scaled by $\delta$, $l/\delta$ becomes nearly invariant with $h^+$, confirming that the spatial significance of secondary motions is governed by the spanwise characteristic lengthscale of the surface ($S/\delta$) rather than the Reynolds number (\citealp{Vanderwel2015,Stroh2016,Yang2017,Hwang2018,Chung2018,Nikora2019,Wangsawijaya2020,zampino2023scaling,Zhdanov2024}). This finding highlights a dual impact: $i)$ the localised influence of secondary flows near the surface, and $ii)$ their interaction with outer-layer dynamics at larger spanwise wavelengths.

\subsection{A note on wall-normal origin: Global vs Local}
\begin{figure}
\centering
\includegraphics[width=0.45\textwidth]{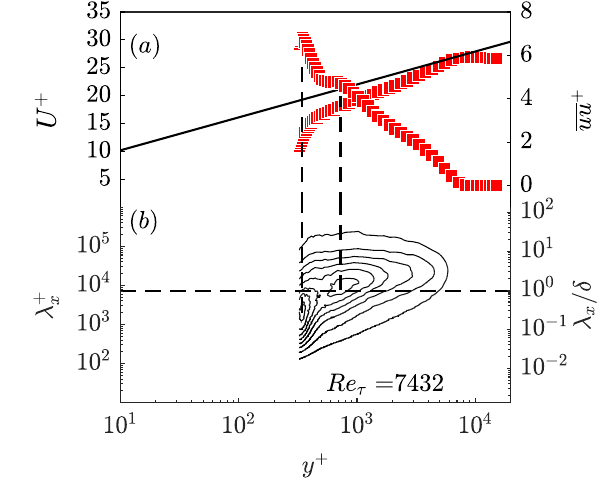}
\includegraphics[width=0.45\textwidth]{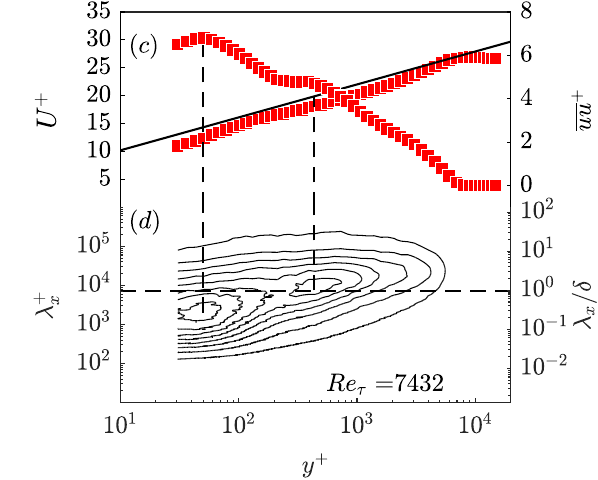}
\caption{$(a, c)$ Wall-normal distributions of the mean streamwise velocity and variance profiles scaled in inner units and $(b, d)$ their associated one-dimensional premultiplied energy spectra, $k_{x}\Phi_{xx}/U_{\tau}^2$. The left panels use a global origin ($y^+_0 = 0$ at the valley), while the right panels use a local origin ($y^+_0 = h^+$ at the ridge tip). Results represent the T100 case at $Re_\tau \approx 7400$. Vertical dashed lines indicate the wall-normal extent of distinct energetic features caused by secondary flows.}
\label{figure5}
\end{figure}

\begin{figure}
\centering
\includegraphics[width=0.32\textwidth]{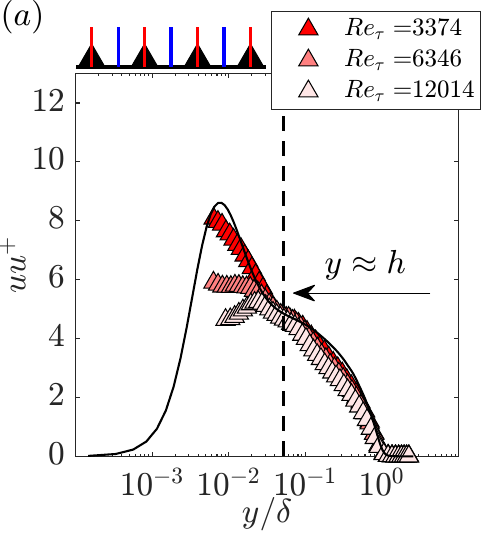}
\includegraphics[width=0.32\textwidth]{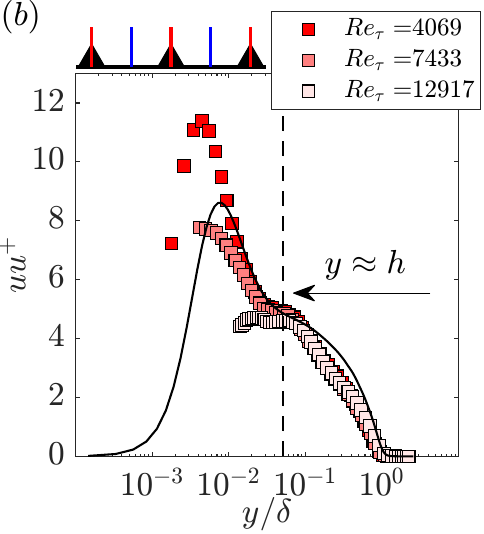}
\includegraphics[width=0.32\textwidth]{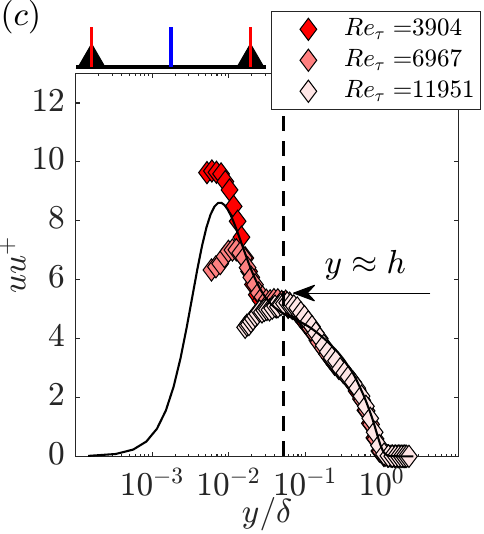}
\includegraphics[width=0.32\textwidth]{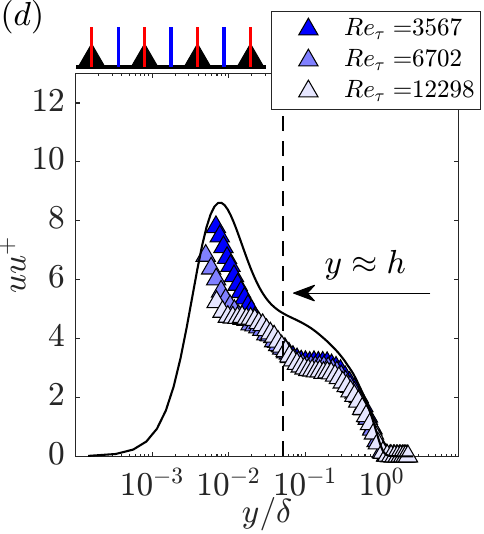}
\includegraphics[width=0.32\textwidth]{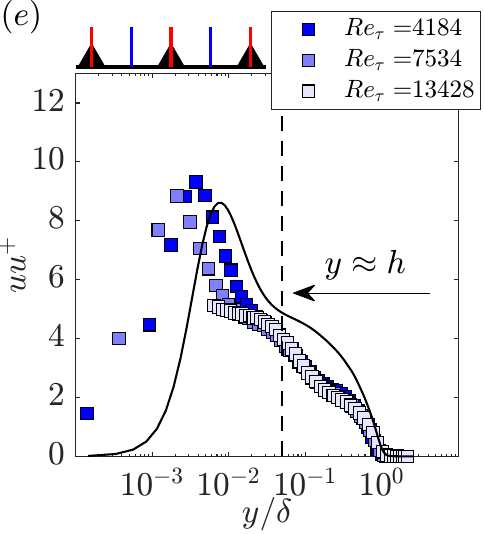}
\includegraphics[width=0.32\textwidth]{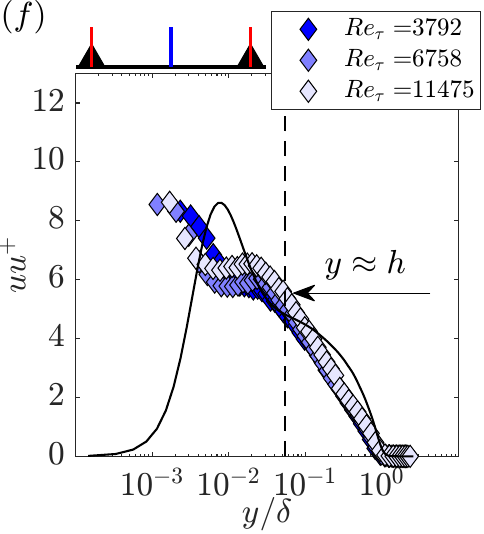}
\caption{Effect of Reynolds number and spanwise spacing on the wall-normal distribution of turbulence intensity profiles, scaled in outer units. Top panels show ridge (LMP) profiles, while bottom panels show valley (HMP) profiles. Increasing Reynolds number is represented by dark to light colour tones. Spanwise spacing increases from left to right panels. Solid black lines depict smooth-wall DNS data from \cite{Sillero2013}.}
\label{figure6}
\end{figure}

Before further examining the impact of surface heterogeneity and secondary flows on turbulence properties, a note on the definition of the wall-normal origin is necessary. Figure~\ref{figure5} illustrates the inner-normalised profiles of the mean streamwise velocity and turbulence intensity, along with a contour map of the premultiplied energy spectra, $k_{x}\Phi_{xx}/U_{\tau}^2$. Taylor's ‘frozen turbulence’ hypothesis was applied to transform the spectra from frequency to wavenumber space, where $k_x = 2\pi/\lambda_x$ represents the streamwise wavenumber, and $\lambda_x = \langle U \rangle / f$ is its associated wavelength, with $f$ being the frequency. The choice of $\langle U \rangle$ as the convection velocity introduces some uncertainty into the estimated length scales, as the true convection velocity is scale-dependent \citep{dennis2008limitations, squire2017applicability}. Nevertheless, using a constant convection velocity is consistent with previous experimental studies \citep{Hutchins2007,monty2009comparison,Squire2016}, enabling direct comparisons with earlier findings. The results in panels $(a, b)$ are scaled using a global origin ($y^+ = 0$ at the valley), while panels $(c, d)$ use a local origin ($y^+ = 0$ at the ridge tip). Although the data in the left and right panels of figure~\ref{figure5} are identical, the choice of wall-normal origin significantly affects how near-wall changes caused by surface heterogeneity are visualised. Using a global origin blends the contributions from near-wall and outer regions, making it harder to isolate the specific effects of secondary flows near the ridge symmetry plane. In contrast, the local origin aligns the near-wall region, enabling a more precise assessment of turbulence redistribution and energy spectra near the surface.

From hereon, the local origin will be used in subsequent analyses to more effectively isolate the effects of spanwise heterogeneity, particularly the role of secondary flows in reorganising momentum and turbulence across the boundary layer.

\subsection{Turbulence intensity}\label{sec:TurbInt}

The influence of spanwise wavelength ($S/\delta$) and Reynolds number on turbulence intensity is examined at both the LMP and HMP locations. Figure~\ref{figure6} presents the inner-normalised turbulence variance profiles as a function of the outer-scaled wall-normal distance, with the top row depicting ridge (LMP) profiles and the bottom row illustrating valley (HMP) profiles. These results reveal distinctive characteristics in all variance profiles induced by surface heterogeneity.

Above the ridges (LMP), turbulence intensity in the near-wall region decreases as the Reynolds number increases. While this trend contrasts with the classical behaviour of homogeneous near-wall turbulence, where intensity typically increases with Reynolds number \citep{Squire2016}, it is consistent with previous findings on heterogeneous surfaces \citep{Medjnoun2018}. This suggests a different mechanism governing large-to-small scale interactions, and how large-scale outer structures modulate near-wall turbulence in the HMPs and LMPs \citep{Pathikonda2017}. At higher $Re_\tau$, an additional energetic peak emerges further from the ridge tip, around $y \approx 2h$, indicating secondary flow interactions that redistribute energy from near-wall turbulence to regions further from the surface. Beyond this height, the profiles collapse and become Reynolds-number-independent, suggesting a universal behaviour in the outer layer.

Above the valleys (HMP), similar trends are observed, with turbulence intensity in the near-wall region decreasing as $Re_\tau$ increases. The profiles collapse around $y \approx h$ for all cases, though the response differs from the ridge profiles. At higher $Re_\tau$, turbulence above the ridges exhibits a single energetic peak for all spanwise wavelengths, whereas above the valleys, a double plateau appears for the T50 and T100 cases ($S/\delta < 1$), reflecting enhanced modulation by secondary flows. For the T200 case ($S/\delta > 1$), this double plateau transitions to a single energetic peak followed by a logarithmic decay, suggesting a different influence of secondary flow at larger spanwise wavelengths.

These observations highlight the interplay between Reynolds number and spanwise wavelength in modifying turbulence intensity. For smaller $S/\delta$, secondary flows more effectively redistribute energy, leading to complex structures such as the double plateau above valleys and stronger ridge-valley (HMP-LMP) interactions. As $S/\delta$ increases, these interactions weaken, and the turbulence intensity profiles converge more rapidly, eventually recovering classical homogeneous behaviour.

\subsection{Spectral characteristics}\label{sec:Spectra}

\begin{figure}
\centering
\includegraphics[width=0.4\textwidth]{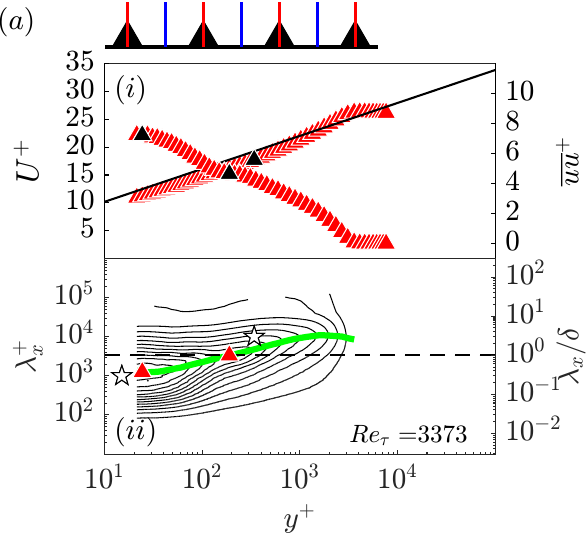}
\includegraphics[width=0.4\textwidth]{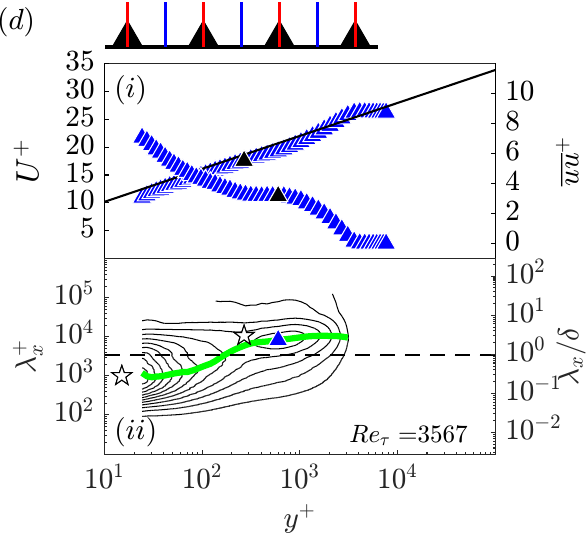}
\includegraphics[width=0.4\textwidth]{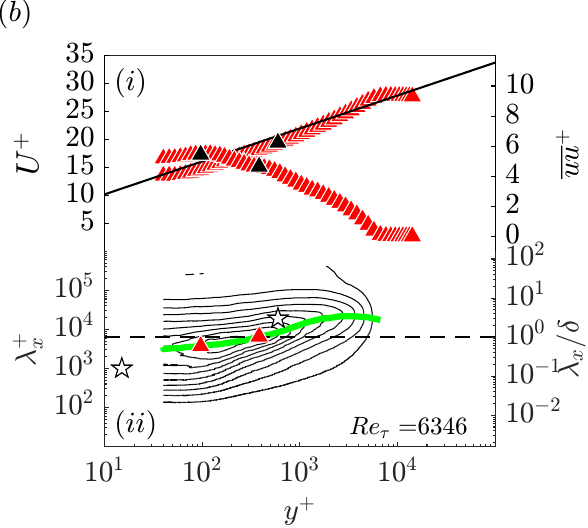}
\includegraphics[width=0.4\textwidth]{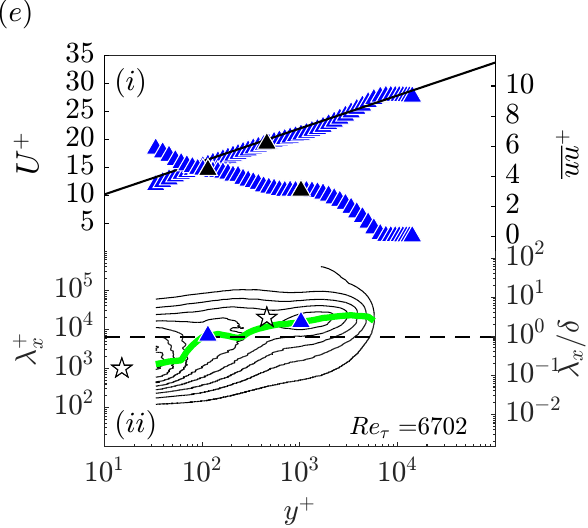}
\includegraphics[width=0.4\textwidth]{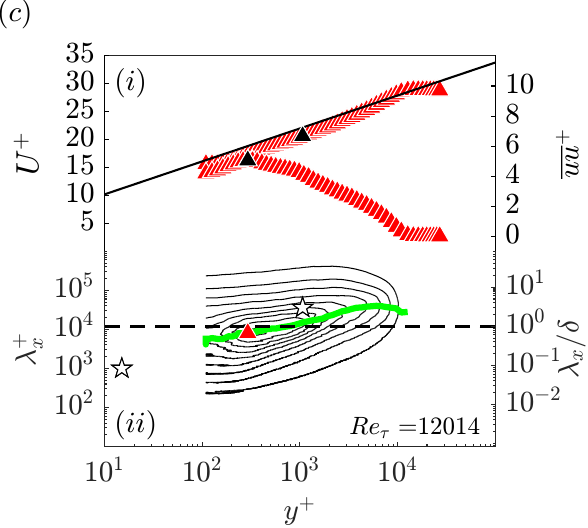}
\includegraphics[width=0.4\textwidth]
{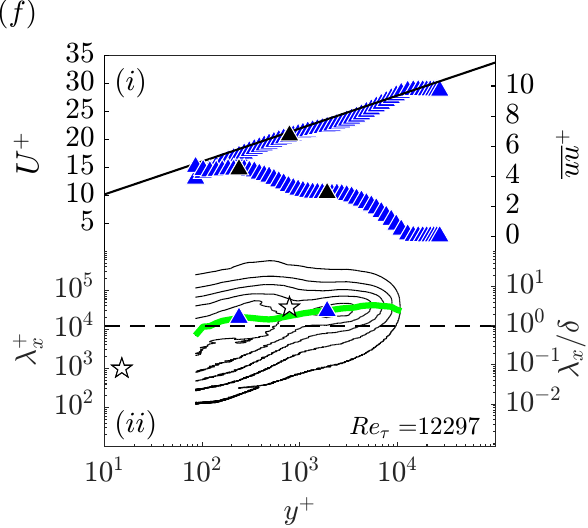}
\caption{$(i)$ Wall-normal distribution of the mean streamwise velocity and variance profiles scaled in inner units, and $(ii)$ their associated one-dimensional premultiplied energy spectrogram, $k_x\Phi_{xx}/U_\tau^2$. The left and right panels show ridge (LMP) in red and valley (HMP) in blue profiles, respectively. From top to bottom, panels represent increasing $Re_\tau$ for the T50 case ($S/\delta \approx 0.3$). The solid line in $(i)$ represents the log-law, while black-filled markers denote the geometric centres of the log-layer and plateau/peaks in the variance profile. Dashed lines in $(ii)$ separate small- and large-scale motions ($\lambda_x = \delta$). The white-filled markers in the $(ii)$ identify the smooth-wall near- and outer-spectral peaks, while colour-filled markers denote new spectral peaks associated with spanwise heterogeneity. The green line represents the local maxima of $k_{x}\Phi_{xx}/U_{\tau}^{2}$ in the $(\lambda_{x}^{+},y^{+})$-plane, illustrating the streamwise coherence of the most energetic structures.}
\label{figure7}
\end{figure}

\begin{figure}
\centering
\includegraphics[width=0.4\textwidth]{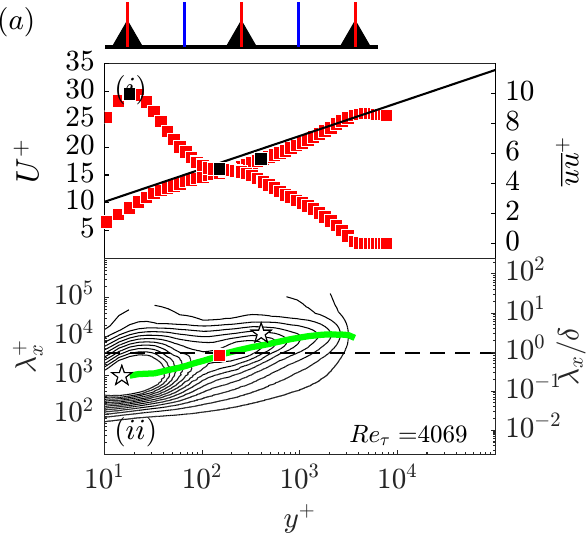}
\includegraphics[width=0.4\textwidth]{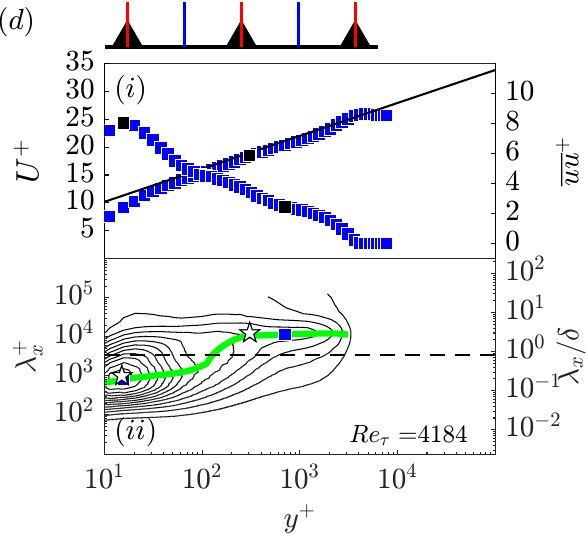}
\includegraphics[width=0.4\textwidth]{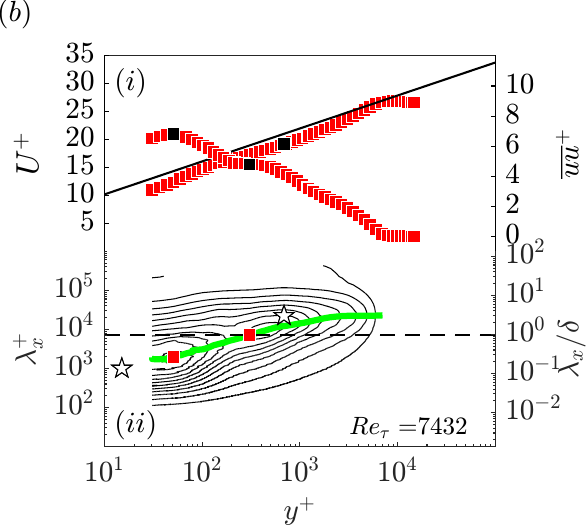}
\includegraphics[width=0.4\textwidth]{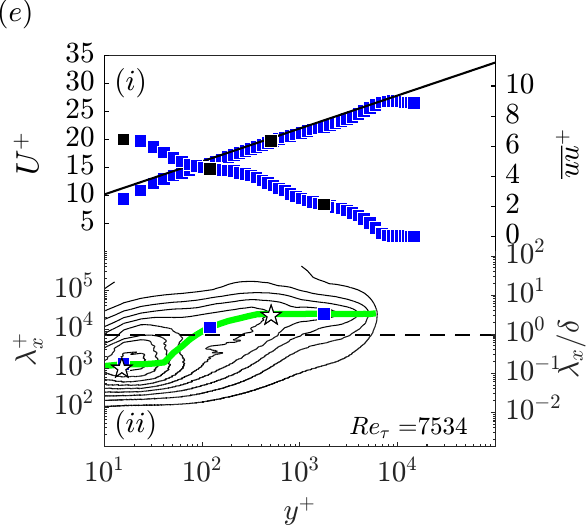}
\includegraphics[width=0.4\textwidth]{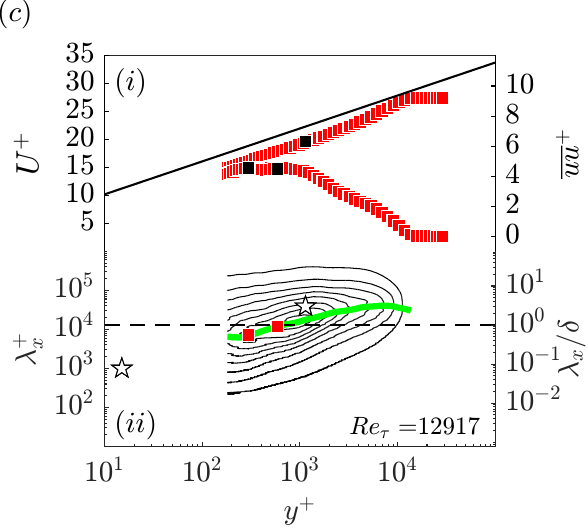}
\includegraphics[width=0.4\textwidth]{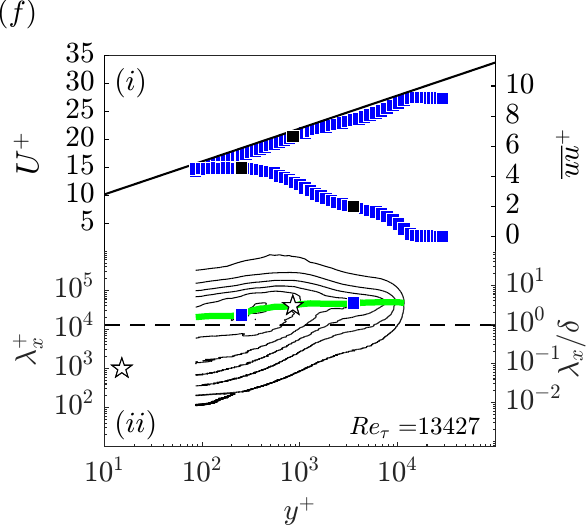}
\caption{Caption in figure \ref{figure7}. Case T100 ($S/\delta \approx 0.6$).}
\label{figure8}
\end{figure}

\begin{figure}
\centering
\includegraphics[width=0.4\textwidth]{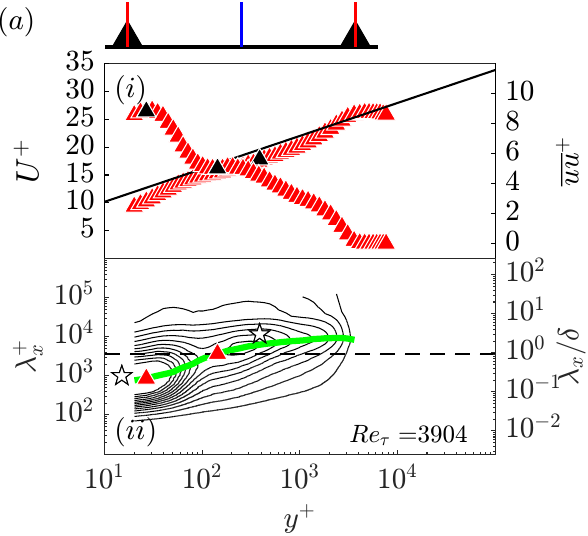}
\includegraphics[width=0.4\textwidth]{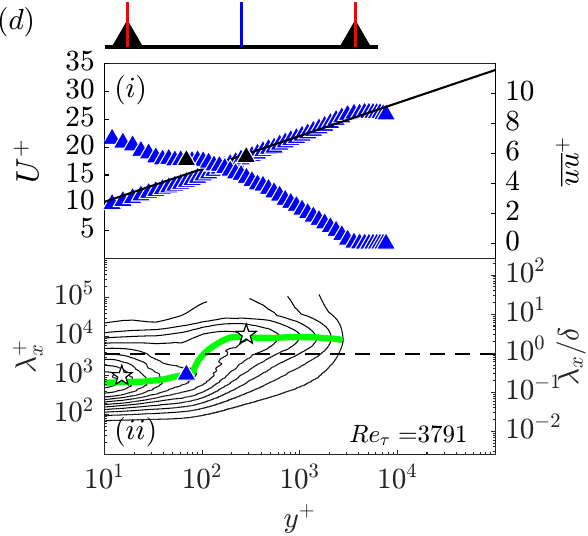}
\includegraphics[width=0.4\textwidth]{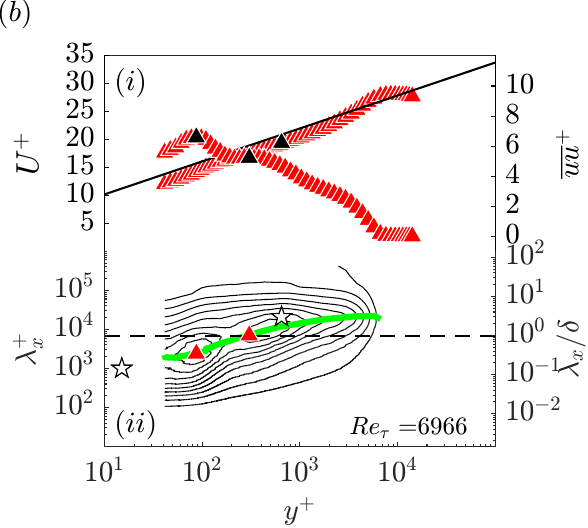}
\includegraphics[width=0.4\textwidth]{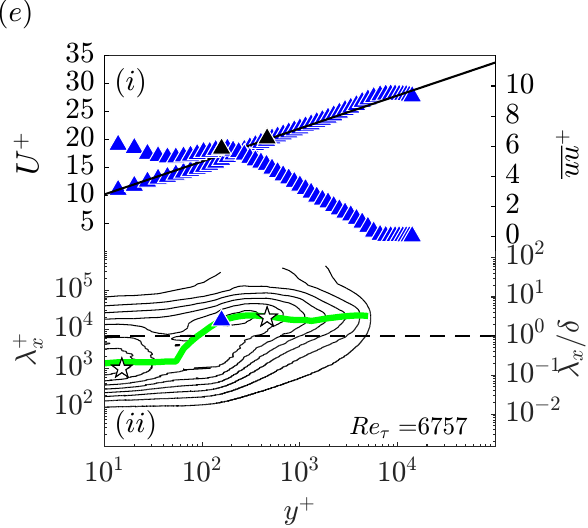}
\includegraphics[width=0.4\textwidth]{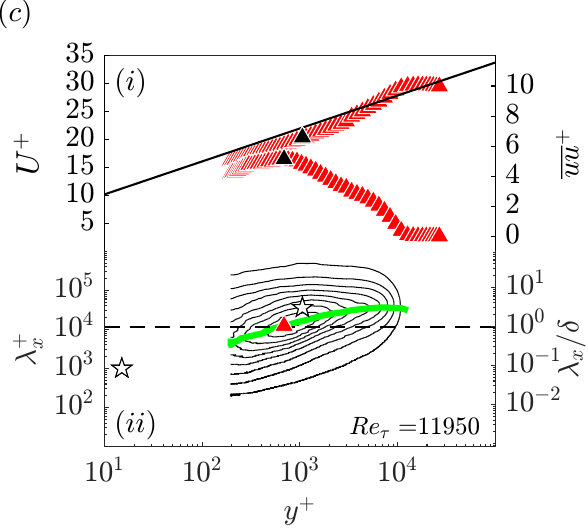}
\includegraphics[width=0.4\textwidth]{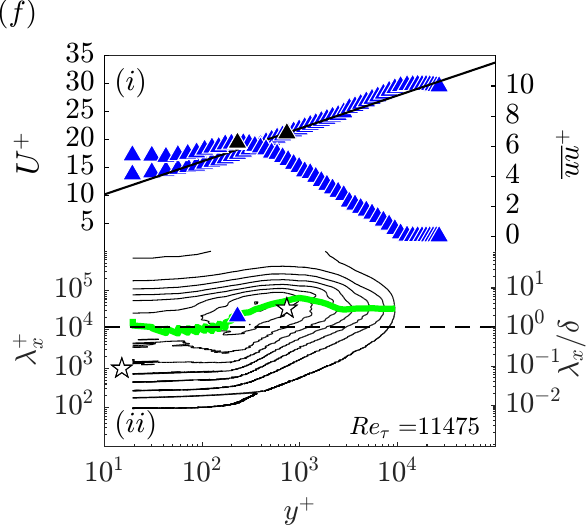}
\caption{Caption in figure \ref{figure7}. Case T200 ($S/\delta \approx 1.3$).}
\label{figure9}
\end{figure}

While the variance profiles provide valuable insights into the local distribution of turbulence structures, they do not fully capture how spanwise heterogeneity and secondary flows influence turbulence across specific scales. To address this, we examine the inner-normalised premultiplied energy spectra of streamwise velocity fluctuations, plotted as functions of wall-normal distance, $y^{+}$, and streamwise wavelength, $\lambda_{x}^{+}$. These spectra, combined with mean velocity and variance profiles, provide a detailed view of turbulence dynamics driven by spanwise heterogeneity and secondary flows.

Figure~\ref{figure7} presents results for the T50 case ($S/\delta \approx 0.3$), with the left and right panels representing the ridge and valley symmetry planes, respectively. Each figure comprises ($i$) the mean and variance profiles, and ($ii$) the associated premultiplied energy spectrograms. The panels demonstrate the effect of increasing Reynolds number, with $Re_{\tau}$ ranging from approximately 3000 to 13000, moving from top to bottom. Key features include the geometric centre of the logarithmic region and the plateau/peaks depicted by the black-filled markers in the velocity and variance profiles, respectively. The smooth-wall near- and outer-spectral peaks, representing the near-wall cycle and VLSM signatures, are shown by the white-filled markers in the spectrograms, while the new spectral peaks associated with spanwise heterogeneity are depicted using coloured markers. The streamwise coherence of the dominant structures across the entire boundary layer is highlighted by the green solid lines.

For T50, the mean velocity profiles on both the ridge and valley symmetry planes exhibit a clear logarithmic distribution when plotted using the local origin. This indicates that, despite the strong surface perturbation imposed on the turbulent boundary layer, the flow remains resilient and retains a logarithmic behaviour similar to that observed over a homogeneous smooth wall \citep{Squire2016}. In this scaling, only a marginal downward shift is observed. However, when using the global origin, a clear momentum loss is evident, reflecting the low-momentum pathways. Conversely, valley profiles show a proportional gain in momentum, offsetting the effects of friction velocity changes. These shifts are indicative of the redistribution of momentum by secondary flows, which transfer high-momentum fluid from the ridge freestream to the valley via upwash and downwash mechanisms. Nevertheless, the combined influence of these shifts is relatively weak, as demonstrated by the roughness function trends shown in figure~\ref{figure2}.

The variance profiles above the ridge plane exhibit a prominent near-wall energetic plateau or peak at $y^{+} \approx 30$, which becomes more pronounced and shifts upward with increasing $Re_{\tau}$. This near-wall peak is distinct from the classical turbulence production peak and indicates significant modulation by large-scale secondary flows. As these secondary flows redistribute energy, a second energetic peak or plateau emerges near the geometric centre of the logarithmic region, becoming more evident in the spectrograms. While this phenomenon has not been previously reported or discussed \textendash primarily due to the lack of studies at high enough Reynolds number \textendash \citet{Medjnoun2018} hinted at similar behaviour in a moderate Reynolds number flow, where a newly observed near-wall spectral peak was shown to shift further from the wall, likely driven by strong upwash motions. This peak nearly merges with the new outer spectral peak, suggesting strong interactions between near-wall turbulence structures and $\delta$-scale structures. The resulting dual-peaked structure highlights the amplified coherence between small- and large-scale motions in regions influenced by secondary flows.

Conversely, in the valley symmetry-plane, turbulence intensity is notably weaker in the logarithmic region. However, a clear and distinctive outer-layer peak emerges above the geometric centre of the logarithmic region, even at low $Re_{\tau}$. As Reynolds number increases, an additional and distinct plateau forms closer to the wall at $y^{+} \approx 300$, suggesting a more localised energy redistribution. Observing the associated spectrograms reveals further details: the outer-layer turbulence peak is associated with an outer-layer spectral peak at approximately $\lambda_{x} \approx 3\delta$, which remains present irrespective of Reynolds number, hence, representing the energetic footprint of the turbulent secondary flow structures. As Reynolds number increases, a new spectral peak emerges from the near-wall region but remains confined below the geometric centre of the logarithmic region, exhibiting a $\delta$-scale streamwise coherence.

The asymmetry between the ridge (LMP) and valley (HMP) profiles is a hallmark of secondary flows. While the LMP exhibits stronger near-wall turbulence modulation and elevated energy levels in the logarithmic region, the HMP shows a weakening of turbulence intensity and a delayed emergence of secondary energetic plateaus. This redistribution of energy across the ridge and valley regions highlights how secondary flows fundamentally alter turbulence dynamics in spanwise heterogeneous flows by modifying the LSMs and completely suppressing the VLSMs.

Building on the T50 case shown in figure~\ref{figure7}, figures~\ref{figure8} and~\ref{figure9} explore the trends for T100 ($S/\delta \approx 0.6$) and T200 ($S/\delta \approx 1.3$), respectively. For T100 and T200, the variance profiles and energy spectra reveal similarly contrasting behaviours between the HMPs and LMPs. At the LMP, for both cases, the variance profiles exhibit trends akin to T50, showing both a near- and outer-layer turbulence peak. These peaks are associated with a modified near-wall cycle that is pushed farther from the wall and begins merging with the outer peak ($\lambda_{x} \approx \delta$).

Conversely, the HMP profiles show larger modifications across the entire profile, with the outer-layer peak retaining a narrow-band wavelength centred at ($\lambda_{x} \approx 3\delta$). As $S/\delta$ increases, secondary flow effects diminish. However, with increasing Reynolds number, a spectral peak near the geometric centre of the logarithmic region becomes more prominent, indicating the reemergence of VLSMs, albeit at lower wavelengths with ($\lambda_{x} \approx 2\delta$). Compared to T50, the near-wall energetic peak is less pronounced, likely due to the influence of downwash motions at the HMP. Additionally, the spectral peaks in the logarithmic region are less distorted. Nonetheless, secondary flows still exert a moderate influence on energy distribution, as evidenced by the presence of secondary energetic plateaus and modulations in the coherence of large-scale motions.

For T200, where $S/\delta \geq 1$, the secondary flows increase and reach their maximum spatial extent but equally have reduced influence on the coherent structures in the HMP. Turbulence dynamics gradually converge toward a more homogeneous turbulent boundary layer behaviour in the valley. While the very near-wall region is unresolved, the variance profiles exhibit a single peak below the geometric centre of the logarithmic region, and the energy spectra begin to recover towards smooth-wall behaviour. At the highest Reynolds number, VLSMs start to become prominent in the logarithmic region ($\lambda_{x} \sim 3-6\delta$), and the outer-layer structures display a behaviour consistent with a canonical turbulent boundary layer. This progression reflects the diminishing influence of spanwise heterogeneity as $S/\delta$ increases beyond unity, with secondary flows becoming less influential in the HMPs as opposed to the LMPs.

The results highlight a consistent trend in turbulence dynamics across the three configurations. For smaller $S/\delta$ (T50), secondary flows are spatially confined yet play a dominant role in redistributing energy, reducing the coherence of large-scale structures, and introducing new spectral peaks in both high- and low-momentum pathways. At intermediate $S/\delta$ (T100), secondary flows grow in strength and continue to modulate turbulence in the LMPs, while their influence in the HMPs weakens. This results in more localised energy redistribution and increased spanwise heterogeneity. At larger $S/\delta$ (T200), secondary flows reach their maximum wall-normal extent but exert minimal influence on the HMPs, whilst strongly modulating the LMPs. This configuration produces the greatest disparity in turbulence behaviour across the span, highlighting the asymmetric impact of secondary motions.

Across all cases, increasing $Re_{\tau}$ further amplifies the influence of secondary flows in the LMPs. These flows consistently modulate the near-wall cycle, impact large-scale motions, and suppress VLSMs, suggesting a degree of universality in their role. In contrast, the HMPs exhibit Reynolds-number-dependent behaviour. As $S/\delta$ increases, secondary flows become more confined, and scale separation between inner and outer motions increases, promoting the recovery of VLSMs in the HMPs. At large $S/\delta$ and high $Re_{\tau}$, turbulence transitions toward behaviour more characteristic of canonical wall-bounded flows, particularly in the high-momentum regions. This evolution underscores the interplay between spanwise heterogeneity and Reynolds number in regulating energy distribution across the boundary layer.

\subsection{Effect of secondary flows on coherent structures}\label{sec:VLSMs}

\begin{figure}
\centering
\includegraphics[width=0.45\textwidth]{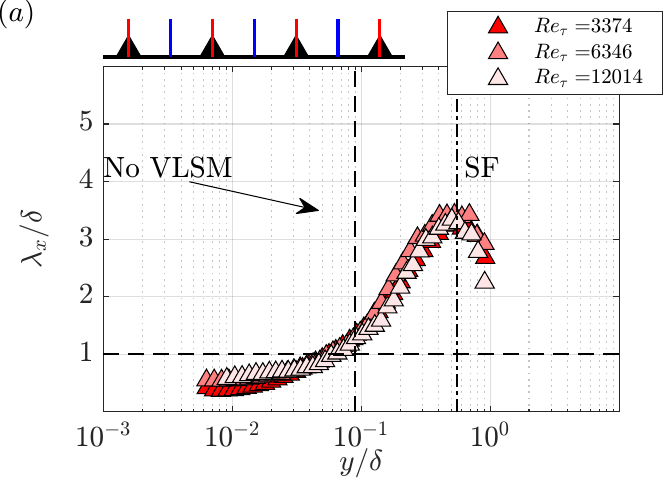}
\includegraphics[width=0.45\textwidth]{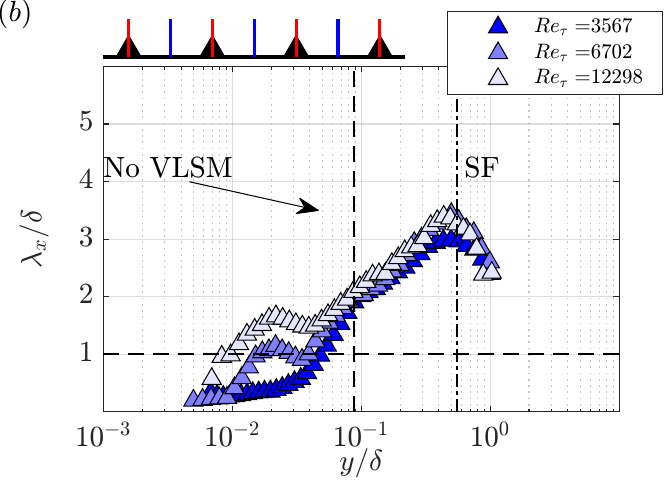}
\includegraphics[width=0.45\textwidth]{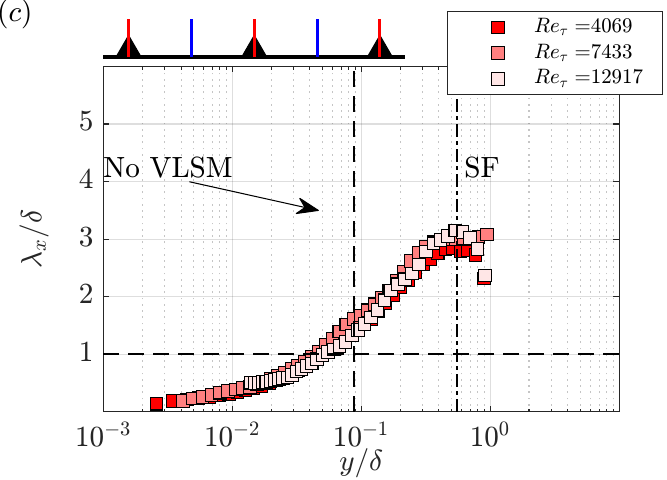}
\includegraphics[width=0.45\textwidth]{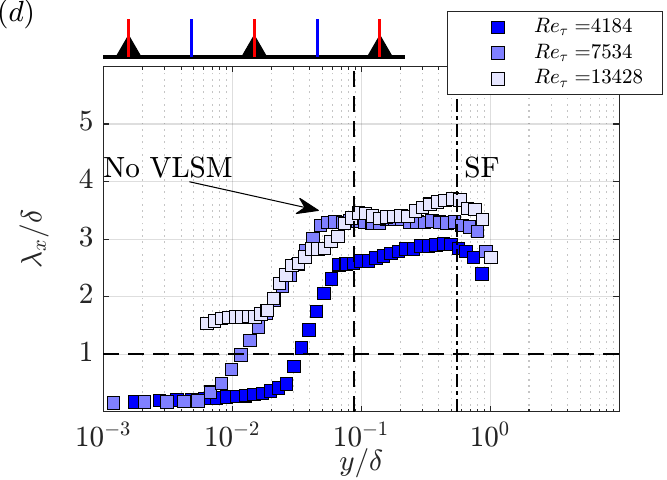}
\includegraphics[width=0.45\textwidth]{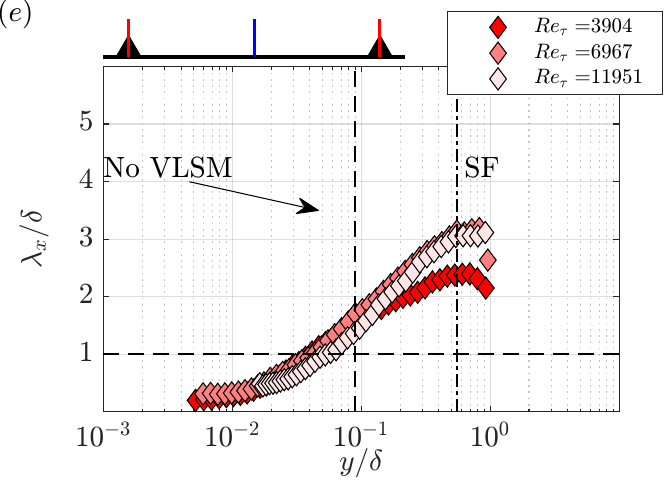}
\includegraphics[width=0.45\textwidth]{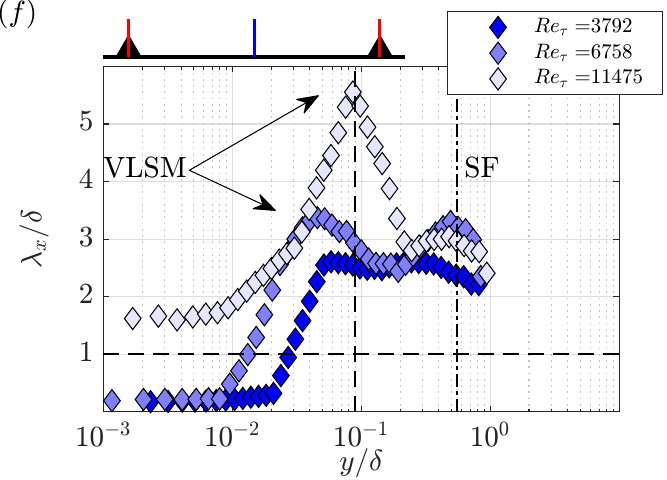}
\caption{Effect of the spanwise spacing and Reynolds number on the streamwise coherence $\lambda_{x}/\delta$ of the turbulence structures across the turbulent boundary-layer, with the vertical dashed-line representing the geometric centre of the logarithmic region where VLSMs are expected to be strongest, whereas the dotted-dashed-line representing the wall-normal height where the secondary flow structures are most prominent. Top to bottom panels represent increase in spanswise spacing whereas left and right panels depict the ridge and valley profiles, respectively.}
\label{figure10}
\end{figure}

\begin{figure}
\centering
\includegraphics[width=0.45\textwidth]{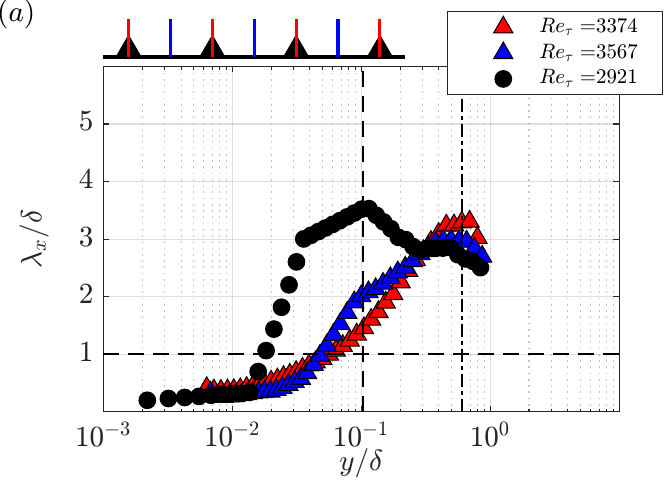}
\includegraphics[width=0.45\textwidth]{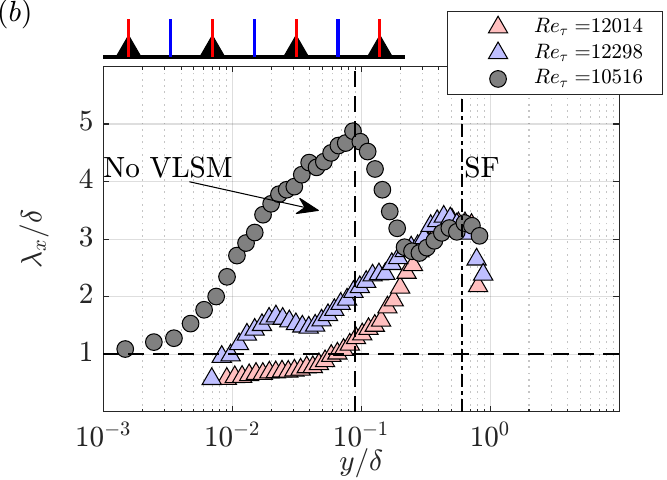}
\includegraphics[width=0.45\textwidth]{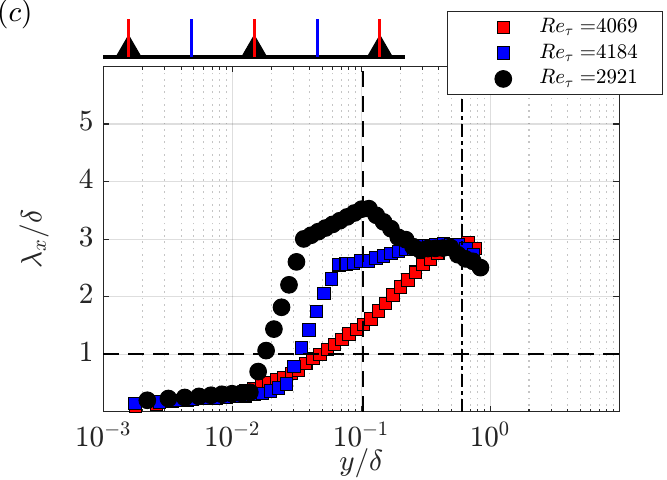}
\includegraphics[width=0.45\textwidth]{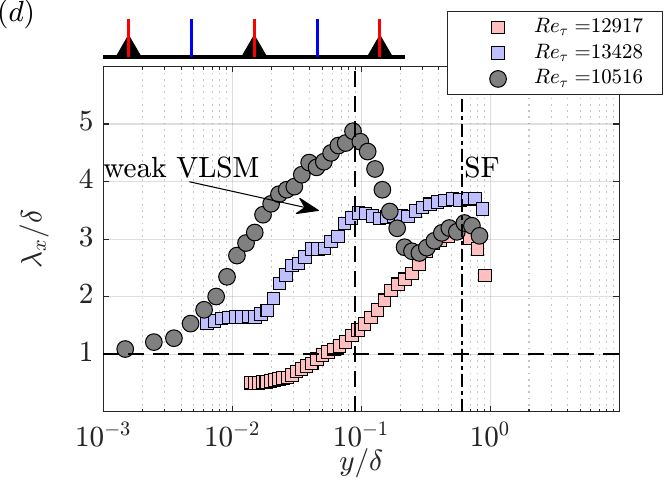}
\includegraphics[width=0.45\textwidth]{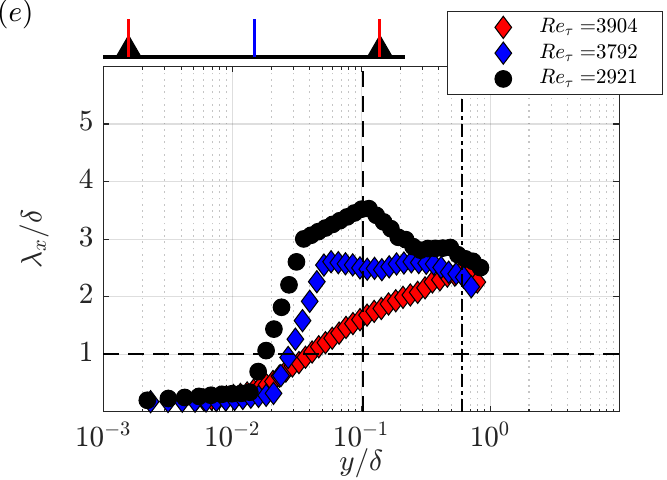}
\includegraphics[width=0.45\textwidth]{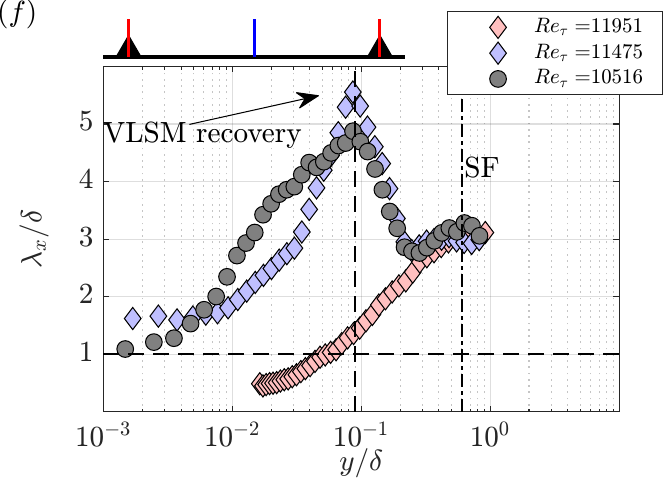}
\caption{Same as in figure \ref{figure10}, but comparing ridge versus valley profiles in each figure, with left and right panels depicting low to high Reynolds number variation.}
\label{figure11}
\end{figure}

Figures~\ref{figure10} and~\ref{figure11} provide a quantitative assessment of how secondary flows influence the suppression, modulation, and recovery of VLSMs across the ridge (LMP) and valley (HMP) symmetry planes. These figures examine the effects of spanwise spacing and Reynolds number on the streamwise coherence $\lambda_{x}/\delta$ of turbulence structures throughout the boundary layer, emphasising the interplay between naturally occurring coherent structures and secondary flow-induced dynamics. The progression from strong but localised secondary flow effects at \(S/\delta = 0.3\) (T50) to broader but weaker effects at \(S/\delta = 1.3\) (T200) underscores the pivotal role of spanwise heterogeneity in modulating turbulence structures.

The left panel of figure~\ref{figure10} shows that, irrespective of spanwise spacing, secondary flows significantly disrupt VLSMs above the ridge symmetry plane. Upwash motions redistribute energy from large scales (\(\lambda_x > \delta\)) to smaller scales, leading to suppressed coherence of large structures. This suppression is evident in the premultiplied energy spectra, where the characteristic spectral peaks in the logarithmic region are absent. However, a consistent, dominant peak emerges across all Reynolds numbers and spanwise spacings at \(y \approx 0.5\delta\) with a streamwise coherence of \(\lambda_{x} \approx 3\delta\). This universal signature of secondary flows persists across configurations, indicating a structural consistency that is independent of Reynolds number and spanwise spacings.

In contrast, the right panel of figure~\ref{figure10} shows substantial variation as the spanwise spacing increases. For \(S/\delta = 0.3\) (T50), VLSM signatures are nearly absent above the valley symmetry plane, where HMPs experience intense suppression. A small but weak coherent signature emerges at the lower boundary of the logarithmic region, becoming more pronounced with increasing Reynolds number. Similar to the ridge plane, a dominant secondary flow peak persists at \(y \approx 0.5\delta\) with \(\lambda_{x} \approx 3\delta\), highlighting comparable outer-layer turbulence dynamics between the LMPs and HMPs.

At intermediate spanwise spacing (\(S/\delta = 0.6\), T100), secondary flow influence diminishes, and turbulence transitions towards a more balanced state. At the HMP, VLSMs begin to recover alignment at the geometric centre of the logarithmic region, with spectral peaks regaining coherence and energy. On the other hand, secondary flow effects weaken but remain evident, particularly in the outer layer. This transitional configuration bridges the strong modulation observed in T50 and the near-recovery seen in T200.

For \(S/\delta = 1.3\) (T200), secondary flows have minimal impact on turbulence dynamics at HMPs, where VLSM signatures fully recover even at intermediate Reynolds numbers. The spectral characteristics align closely with homogeneous smooth-wall behaviour. Although secondary flows remain observable at the valley plane, their prominence is significantly reduced, allowing VLSMs to dominate the energy spectrum. This recovery reflects the gradual transition to homogeneity as spanwise spacing increases, with turbulence dynamics increasingly governed by classical wall-bounded mechanisms.

The combined analysis of figures~\ref{figure10} and~\ref{figure11} highlights a critical threshold around \(S/\delta \sim 1\), beyond which the influence of spanwise heterogeneity diminishes significantly. At smaller \(S/\delta\), secondary flows strongly disrupt the energy cascade, suppressing VLSMs and enhancing interactions between near-wall small-scale motions and outer-layer large-scale motions, particularly above the ridge symmetry plane. Meanwhile, the LMP remains largely unaffected, exhibiting a universal behaviour that is independent of surface heterogeneity and Reynolds number. In contrast, the effect of surface heterogeneity on the HMP weakens as \(S/\delta\) increases. This weakening enables large-scale motions to recover coherence while coexisting with secondary flow structures. Once \(S/\delta\) exceeds unity, turbulence dynamics gradually recover and closely resemble those of homogeneous smooth-wall flows, marking a significant transition.

\begin{figure}
\centering
\includegraphics[height=0.25\textwidth]{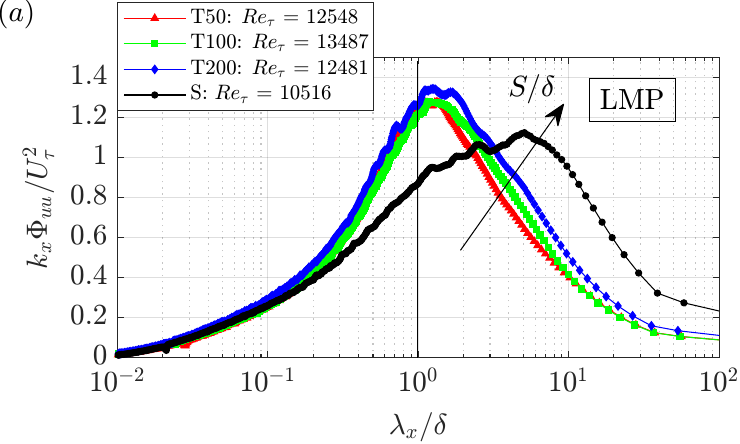}
\includegraphics[height=0.25\textwidth]{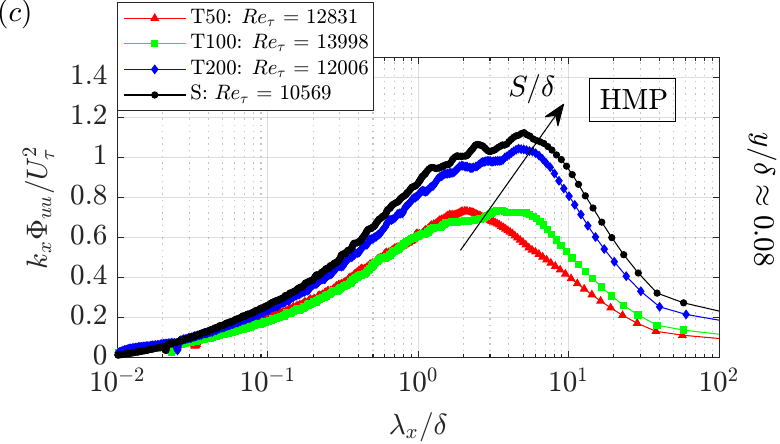}
\includegraphics[height=0.25\textwidth]{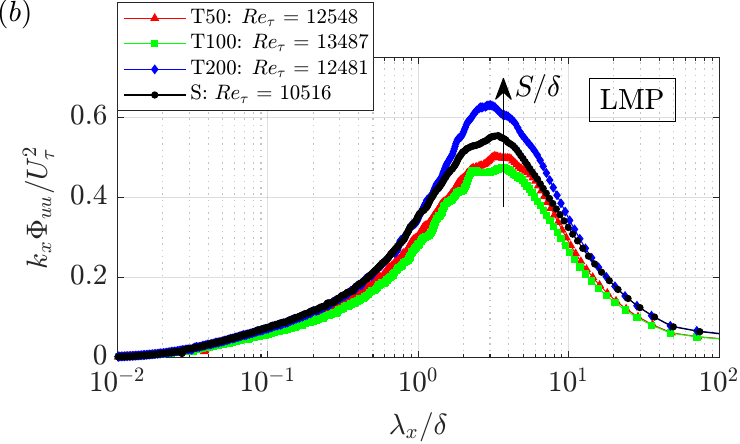}
\includegraphics[height=0.25\textwidth]{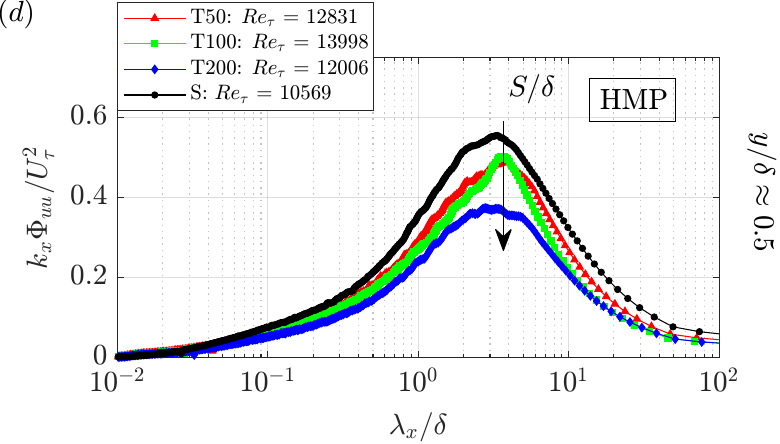}
\caption{Effect of the spanwise spacing on the pre-multiplied energy spectra as a function of the outer-normalised wavelength for the (left) ridge and (right) valley symmetry planes, ($a,b$) geometric centre of the logarithmic region where VLSMs are expected to be strongest and ($c,d$) $y/\delta\approx 0.5$ where secondary flow structures are expected to be strongest. The solid black line represents $\lambda_{x}/\delta=1$.}
\label{figure12}
\end{figure}

Figure~\ref{figure12} presents a comparison of the pre-multiplied and outer-normalised streamwise energy spectra, to assess the influence of spanwise spacing ($S/\delta$) on energy distribution across the ridge (left) and valley (right) symmetry planes. Panels $(a,b)$ correspond to the LMP and $(c,d)$ to the HMP, showing results at the geometric centres of the log-layer ($y/\delta \approx 0.15$) and the outer region ($y/\delta \approx 0.5$), where VLSMs and secondary flows, respectively, are most prominent.

Across all $S/\delta$, the ridge profiles exhibit a consistent redistribution of spectral energy driven by upwash motions. In the log-layer, VLSMs ($\lambda_x > 3\delta$) are suppressed and energy is enhanced at intermediate scales ($\lambda_x \sim 1$–$2\delta$), consistent with prior observations \citep{Medjnoun2018,Awasthi2018,Barros2019,Zampiron2020}. This reflects the radial disruption of coherent structures by secondary flows. In all cases, the small-scale energy content ($\lambda_x < \delta$) collapses well, suggesting that secondary flows act primarily on larger scales, whilst smaller scales are passively reorganised.

In contrast, the valley (HMP) spectra display greater sensitivity to $S/\delta$. At low $S/\delta = 0.3$, significant VLSM suppression is observed, while at larger $S/\delta = 1.3$, the spectra more closely resemble the smooth-wall baseline. This indicates a partial recovery of classical boundary layer dynamics in the HMPs, although residual suppression of VLSMs persists, likely due to spanwise and wall-normal momentum exchanges imposed by secondary flow structures.

A persistent outer-layer spectral peak at $\lambda_x \approx 3\delta$, $y/\delta \approx 0.5$, is observed across all cases and configurations. Its intensity increases with $S/\delta$ in the LMPs but decreases in the HMPs, marking a robust spectral footprint of secondary flows. The upwash-dominated ridge plane consistently exhibits stronger modulation of large-scale energy, while the valley plane shows reduced but detectable influence. This universal peak provides compelling evidence of secondary flow organisation across the boundary layer, even as the intensity and spanwise asymmetry evolve with increasing $S/\delta$.

Overall, the ridge–valley comparisons reveal distinct but complementary roles: the ridge consistently shows enhanced intermediate-scale energy via upwash-induced redistribution, while the valley reflects a transitional zone where VLSMs may be suppressed or recovered depending on $S/\delta$. This asymmetry highlights the nuanced influence of secondary flows and the central role of spanwise heterogeneity in shaping turbulent energy spectra.

\section{Discussion}

The presence of spanwise surface heterogeneity in wall-bounded turbulent flows has long been recognised as a driver of secondary motions, which in turn influence the dynamics of large-scale coherent structures. Foundational works \citep{Hinze1973, Anderson2015} have shown that streamwise-aligned heterogeneities in surface roughness or elevation generate secondary flows of the second kind, arising from spatial imbalances in turbulence production and dissipation. These secondary motions organise the flow into high- and low-momentum pathways, whose topology and strength are primarily governed by the characteristic spanwise heterogeneity length scale, typically expressed as $S/\delta$.

A number of key questions have guided the progression of research in this area: $i$) how should surface heterogeneity be defined, and what parameters such as heterogeneity type, geometry, and relevant length scales govern the onset and spatial extent of secondary flows? $ii$) what is the nature of their interaction with large-scale structures, particularly very-large-scale motions? $iii$) to what extent are these effects dependent on Reynolds number? and $iv$) how should turbulence be scaled or modelled in the presence of strong heterogeneity? Building on previous work, the current study bridges some of these gaps by examining the spectral and structural evolution of secondary flows and their modulation of turbulence at multiple scales. The use of a wide Reynolds number range ($Re_\tau \approx 3\,000$--$13\,000$) enables a clearer decoupling of inner- and outer-layer dynamics, facilitating a systematic evaluation of scale interactions.

In agreement with regime maps established in the literature (see for example \citealt{Willingham2014,Vanderwel2015,Yang2017,Chan2018,Chung2018}), we observe that mean and turbulent flow heterogeneity are most prominent for $S/\delta \sim \mathcal{O}(1)$ and exhibit differing degrees of wall-normal influence depending on the spanwise spacing. For small spacings ($S/\delta < 1$), secondary flows remain confined near the wall and exert strong modulation of the near-wall turbulence cycle. As $S/\delta$ increases, these motions become more vertically extended, although their influence on wall shear stress and turbulence intensity diminishes in the HMPs. This transition is reflected in both drag measurements and turbulence statistics. These results collectively indicate that spanwise heterogeneity reorganises the turbulence cascade and scale interactions via secondary flow structures.

The present study also provides new clarity regarding the interaction between secondary flows and VLSMs. Previous experimental and numerical studies have reported seemingly conflicting conclusions: \citet{Zampiron2020} and \citet{Luo2023} observed suppression of VLSMs in the presence of secondary motions, whereas \citet{Wangsawijaya2020} and \citet{Barros2019} documented partial recovery or coexistence under certain conditions. Our results reconcile these findings by demonstrating that suppression and coexistence are not mutually exclusive outcomes but instead depend on spanwise spacing ($S/\delta$), wall-normal location, and momentum pathway. Specifically, VLSMs are consistently suppressed in the LMPs, regardless of $Re_\tau$ and $S/\delta$, but begin to recover in the HMPs as $S/\delta \gtrsim 1$ and Reynolds number increases. Moreover, a quasi-universal spectral peak emerges at $\lambda_x \approx 3\delta$, $y/\delta \approx 0.5$, across all configurations tested. This peak remains visible even when classical VLSMs reappear, suggesting that it represents a robust spectral footprint of energy redistribution associated with secondary flow structures. The data also reveal a pronounced asymmetry: turbulence intensities in the LMPs increase with $Re_\tau$ in both the near-wall and logarithmic regions, while the HMPs exhibit attenuated turbulence and a gradual return to canonical boundary layer behaviour.

These findings align well with earlier studies. \citet{Awasthi2018} identified a persistent spectral peak termed ``modulators'' in the LMP but not in the HMP, consistent with our observations. Similarly, \citet{Barros2019} reported a merging of the LSM and VLSM peaks into a single spectral mode with $\lambda_x \approx 3\delta$ along the LMP, while little modulation was observed in the HMP. \citet{Zampiron2020} further confirmed that VLSMs are strongly suppressed at small $S/\delta$, with gradual recovery for $S/\delta \gtrsim 1$, again in agreement with our results.

In contrast, \citet{Wangsawijaya2020} observed that secondary flows and LSM/VLSMs can coexist in the asymptotic limits of $S/\delta \ll 1$ and $S/\delta \gg 1$. While this may initially appear inconsistent with the present findings, the discrepancy is reconcilable when differences in surface topology are considered. Their study employed strip-type heterogeneity with no elevation variation, whereas the current work uses ridge-type surfaces with finite wall-normal height. We hypothesise that in strip-type configurations, the absence of vertical forcing permits the re-emergence of VLSMs at small $S/\delta$. In contrast, ridge-type surfaces generate finite-height-induced secondary motions that persist even at small $S/\delta$, remaining strong enough to disrupt VLSM formation. These results suggest that the interaction between secondary flows and large-scale structures is governed not only by $S/\delta$ but also by the surface type and the physical mechanism responsible for generating secondary motions.

These insights highlight the central role of secondary flows in shaping turbulence dynamics and offer clear implications for engineering design. At small $S/\delta$, secondary motions enhance mixing by disrupting the classical energy cascade and intensifying interactions across scales, desirable features in applications involving scalar transport or surface renewal. In contrast, larger $S/\delta$ configurations preserve canonical turbulence behaviour in certain regions, particularly the HMPs, while still enabling modulation in the LMPs. Such configurations are advantageous in scenarios requiring drag reduction or improved aerodynamic efficiency, where a balance between coherence and local mixing is needed.

Despite these advances, several open questions remain. First, the origin and dynamics of secondary flow meandering as suggested by \citet{Wangsawijaya2022} and \citet{Luo2023} remain not fully understood. The observed spectral peak at $\lambda_x \approx 3\delta$ may be linked to such meandering, but further spatio-temporal measurements are required to establish causality. Second, the ``universality'' of these findings to rough or multiscale heterogeneous surfaces remains unclear. Most existing studies, including the present one, employ idealised strip or ridge geometries; real-world surfaces may exhibit different flow organisation mechanisms. Third, the role of inner–outer interaction mechanisms (\citealp{Pathikonda2017, Anderson2018}), in the presence of strong secondary flows remains to be fully characterised. Finally, these findings present new challenges for turbulence modelling. Secondary motions are not adequately captured in traditional RANS frameworks (\citealp{zampino2022linearised,zampino2023scaling,lasagna2024linear}), and new closures may be required to incorporate the energetic and structural influence of spanwise heterogeneity.

In the broader context, this study contributes to the evolving understanding of secondary flows in wall-bounded turbulence and supports the development of a more general framework for flows in heterogeneous environments. The results demonstrate that classical concepts such as outer-layer similarity, spectral scaling, and attached eddy structures must be re-evaluated in regimes where secondary flows dominate. Future work should aim to extend these insights towards practical flow control strategies, including the design of engineered surfaces that harness secondary motions for drag reduction, enhanced mixing, or targeted energy redistribution.

\section{Conclusion}

This study has examined the evolution of secondary flows in spanwise-heterogeneous turbulent boundary layers across a wide range of friction Reynolds numbers ($Re_\tau \approx 3\,000$--$13\,000$), with a particular focus on the interaction between secondary motions and naturally-occurring coherent motions. Using direct drag measurements and high-resolution hot-wire anemometry, we demonstrate that the spatial organisation and strength of secondary flow features depend sensitively on $Re_\tau$ and the spanwise characteristic length scale of the surface heterogeneity.

Across all conditions, secondary flows dominate the dynamics above the ridges. These motions draw energy from large-scale structures, suppress the coherence of VLSMs ($\lambda_{x} > 3\delta$), and redistribute energy towards intermediate scales ($\lambda_{x} \sim 1$–$3\delta$), while the small-scale energy content ($\lambda_{x} < \delta$) remains largely invariant. In contrast, the behaviour in the valleys is more sensitive to both Reynolds number and spanwise spacing. At low $Re_\tau$ and small $S/\delta$, secondary motions extend throughout the boundary layer and modulate both the log region and the outer layer. However, as $Re_\tau$ and $S/\delta$ increase, the influence of these motions diminishes in the valleys, and VLSMs begin to re-emerge from the near-wall region. This leads to a complex interaction between re-emerging near-wall VLSMs and residual outer-layer secondary motions, particularly in cases with $S/\delta > 1$, where secondary structures become increasingly localised or reorganised.

This transition reflects a fundamental shift in turbulence structure, wherein inner-layer coherence begins to dominate over surface-induced heterogeneity. The persistence of secondary flow signatures in the outer region, particularly in valleys at high $Re_\tau$ and $S/\delta$, underscores the competing roles of inertia-driven and surface-driven mechanisms in shaping the flow. These findings provide previously undocumented experimental evidence that complements recent studies in heterogeneous wall-bounded turbulence. They also highlight the importance of resolving the Reynolds-dependent balance between large-scale structure organisation and surface-induced secondary motions.

From a practical standpoint, these insights have direct implications for the modelling and control of turbulent flows over structured surfaces. Predictive models that neglect the interplay between heterogeneity and Reynolds number may fail to capture critical features of momentum redistribution and wall stress. In engineering and environmental contexts, this could influence strategies for drag reduction, mixing enhancement, or scalar transport. Future work should explore the threshold $Re_\tau$ values and heterogeneity parameters that govern these transitions, particularly in more complex or multiscale surfaces. The role of spanwise coherence, VLSM footprint evolution, and the potential meandering of secondary flows remains to be fully characterised. A deeper understanding of these phenomena may inform the design of surfaces that selectively exploit or suppress secondary motions, enabling novel passive flow control strategies at high Reynolds numbers.

\backsection[Funding]{We gratefully acknowledge the financial support from EPSRC (Grant ref no: EP/V00199X/1).}

\backsection[Data availability statement]{The data that support the findings of this study will become openly available through UoS repository upon publication}

\backsection[Author ORCIDs]{\\ T.Medjnoun http://orcid.org/0000-0002-8699-1305; \\  B.Ganapathisubramani https://orcid.org/0000-0001-9817-0486. \\ M. Nillson-Takeuchi}

\appendix

\section{}\label{appA}
\begin{figure}
\centering
\includegraphics[width=1\textwidth]{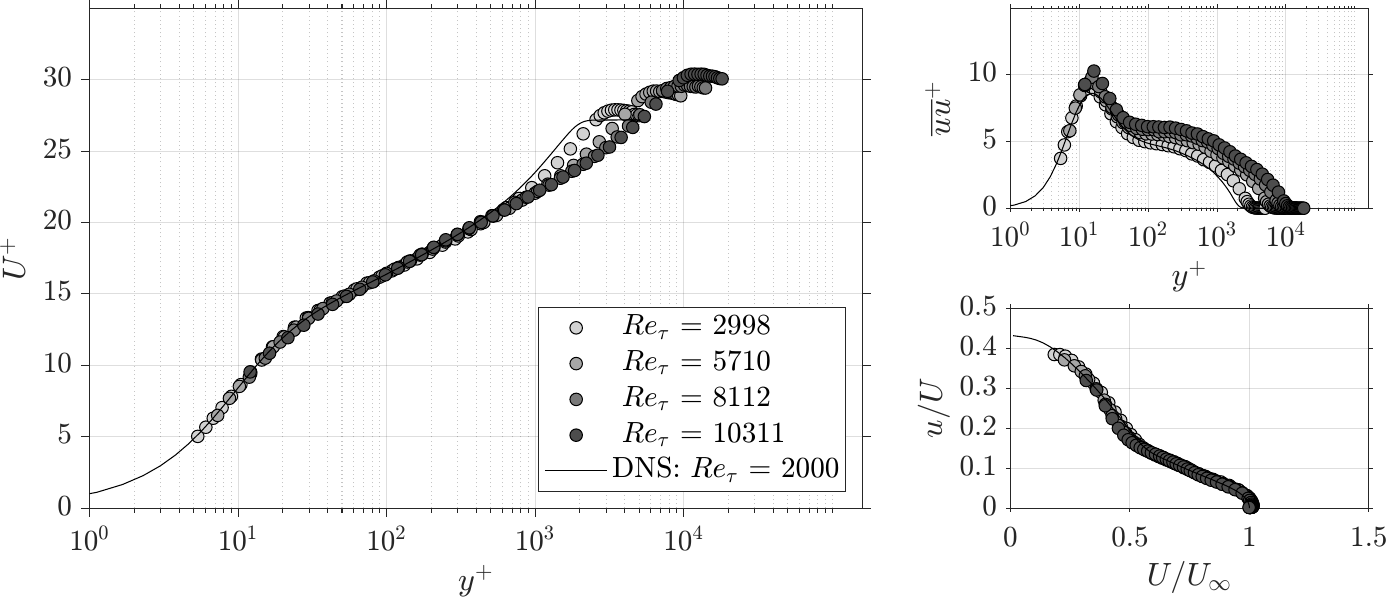}
\caption{Validation results of the Southampton BLWT mean velocity and turbulence intensity profiles. The black solid lines represent DNS data from \cite{Sillero2013}.}
\label{figure13}
\end{figure}

Figure~\ref{figure13} presents smooth-wall validation data obtained from the Southampton Boundary Layer Wind Tunnel (BLWT), comparing experimental measurements of streamwise velocity, turbulence variance, and diagnostic plot results with DNS data from the study of \citet{Sillero2013}. This comparison serves to validate the accuracy and reliability of the experimental setup and measurement techniques used throughout the study.

The streamwise velocity profiles show excellent agreement with the DNS data, represented by the black solid lines, particularly in the logarithmic region of the boundary layer. This alignment confirms that the wind tunnel effectively replicates canonical smooth-wall flow conditions. The turbulence variance profiles also closely match the DNS results, capturing the near-wall peak at $y^+ \approx 15$, which is characteristic of turbulence production in smooth-wall flows. Beyond this region, the variance decays as expected, further demonstrating the fidelity of the experimental setup. The diagnostic plot provides additional validation by highlighting the relationship between turbulence intensity and normalised velocity. The close alignment with DNS trends corroborates the robustness of the measurement techniques in capturing turbulence characteristics such as velocity, variance, and energy spectra.

This validation is crucial for establishing confidence in the current observations. By demonstrating strong agreement with DNS for smooth-wall flows and replicating canonical turbulence behavior, the study ensures that deviations observed in heterogeneous configurations are attributable to spanwise heterogeneity and not experimental artifacts.

\bibliographystyle{jfm}


\end{document}